\pdfoutput=1
\newif\ifarxiv\arxivtrue
\documentclass[conference, a4paper]{IEEEtran}
\usepackage[T1]{fontenc}

\usepackage{amsmath,amssymb,amsfonts,amsthm,amssymb}
\usepackage{cite}
\usepackage{mathtools}
\usepackage{algorithmic}
\usepackage{xcolor, graphicx}
\usepackage{textcomp}
\usepackage{url}
\usepackage[version=4]{mhchem}
\def\BibTeX{{\rm B\kern-.05em{\sc i\kern-.025em b}\kern-.08em
    T\kern-.1667em\lower.7ex\hbox{E}\kern-.125emX}}

\usepackage[pdftex]{hyperref}
\usepackage{orcidlink}

\usepackage{xparse}
\usepackage{bcprules}
\usepackage{braket}
\usepackage{booktabs}
\usepackage{multirow}
\usepackage{siunitx}

\NewDocumentCommand\pipe{}{\mathrel{|}}

\NewDocumentCommand\etal{}{\textit{et al}.\ }

\NewDocumentCommand\defeq{}{\triangleq}

\DeclareMathOperator{\Tr}{Tr}

\def\mycaption{\def\@captype{figure}\caption}

\NewDocumentCommand\mdoubleplus{}{\ensuremath{\mathbin{+\mkern-10mu+}}}

\newtheoremstyle{mystyle}%
  {}
  {}
  {\normalfont}
  {}
  {\bf}
  {}
  {5pt}
  {}
\theoremstyle{mystyle}

\newtheorem{definition}{Definition}

\newtheorem{example}{Example}
\newtheorem{remark}{Remark}

\newif\ifdraftComments
\draftCommentstrue
\def\mkDraftFn#1#2{%
  \expandafter\def\csname #1\endcsname##1{\ifdraftComments\textcolor{#2}{[#1: ##1]}\marginpar[$\longrightarrow$]{$\longleftarrow$}\fi}%
}
\mkDraftFn{SN}{blue}
\mkDraftFn{RW}{red}

\usepackage{bussproofs}

\usepackage{cleveref}
\crefname{figure}{Fig.}{Fig.}
\Crefname{figure}{Fig.}{Fig.}
\crefname{table}{Table}{Table}
\Crefname{table}{Table}{Table}
\crefname{example}{Example}{Example}
\Crefname{example}{Example}{Example}

\usepackage{quantikz} 

\newcommand{\inquirnt}[1]{\mathit{#1}}
\newcommand{\inquirmv}[1]{\mathit{#1}}
\newcommand{\inquirkw}[1]{\mathbf{#1}}
\newcommand{\inquirsym}[1]{#1}

\usepackage{xparse}
\usepackage{bussproofs}
\usepackage{stmaryrd}

\NewDocumentCommand\rstate{mmmmm}{\left[ #1 , #2 , #3, #4, #5\right]}
\NewDocumentCommand\evalarrow{}{\rightarrow}
\NewDocumentCommand\getq{}{\text{get}}

\NewDocumentCommand\ketbra{mmm}{\ket{#1}_{#2}\bra{#3}}
\NewDocumentCommand\hpush{}{\text{push}}

\begin{document}

\title{InQuIR: Intermediate Representation for Interconnected Quantum Computers}

\author{
  \IEEEauthorblockN{Shin Nishio\orcidlink{0000-0003-2659-5930}\IEEEauthorrefmark{1} and Ryo Wakizaka\orcidlink{0000-0001-8762-9335}\IEEEauthorrefmark{2}}
  \IEEEauthorblockA{
    \IEEEauthorrefmark{1}Department of Informatics,
    School of Multidisciplinary Science,\\
    SOKENDAI (The Graduate University for Advanced Studies),
    Tokyo, Japan
  }
  \IEEEauthorblockA{
    \IEEEauthorrefmark{2}Graduate School of Informatics,
    Kyoto University,
    Kyoto, Japan
  }
}

\maketitle
\thispagestyle{plain}
\pagestyle{plain}
\begin{abstract}
Various physical constraints limit the number of qubits that can be implemented in a single quantum processor, and thus it is necessary to connect multiple quantum processors via quantum interconnects. While several compiler implementations for interconnected quantum computers have been proposed, there is no suitable representation as their compilation target. The lack of such representation impairs the reusability of compiled programs and makes it difficult to reason formally about the complicated behavior of distributed quantum programs. We propose InQuIR, an intermediate representation that can express communication and computation on distributed quantum systems. InQuIR has formal semantics that allows us to describe precisely the behaviors of distributed quantum programs. We give examples written in InQuIR to illustrate the problems arising in distributed programs, such as deadlock. We present a roadmap for static verification using type systems to deal with such a problem. We also provide software tools for InQuIR and evaluate the computational costs of quantum circuits under various conditions. Our tools are available at \url{https://github.com/team-InQuIR/InQuIR}.
\end{abstract}

\section{Introduction}
The scale and accuracy of quantum devices have rapidly increased in recent years, leading to small-scale experimental implementations of multiple quantum algorithms\cite{adedoyin2018quantum}, quantum error correction\cite{chiaverini2004realization}, et cetera. In contrast, it turned out that there is a limit to the number of qubits implemented on a single processor due to physical constraints such as the size of the dilution refrigerator\cite{krinner2019engineering}, the silicon wafer \cite{gold2021entanglement} and wiring difficulties\cite{tamate2021toward}. It is necessary to construct a quantum computer cluster by connecting multiple quantum processors with a quantum interconnect\cite{Awschalom2021} to perform large-scale quantum information processing.

Concerning the programs for distributed tasks such as quantum concensus\cite{ben2005fast} and blind quantum computing\cite{barz2012demonstration} on a distributed system, it is easy to explicitly describe the communication between quantum computers required in the quantum information processing. When quantum communication is explicitly described, quantum communication interfaces such as QMPI\cite{hanerDistributedQuantumComputing2021} play an important role. However, the quantum program is not written for distributed systems in most quantum information processing. Also, it is difficult to write a quantum program that satisfies various constraints on distributed systems or to consider strategies to avoid communication bottlenecks.
A program that does not meet such inherent constraints can lead to the failure of the entire computation.

Several compilation methods for distributed quantum computation have been proposed to address such issues and evaluated by execution time\cite{cuomoOptimizedCompilerDistributed2021}, circuit depth\cite{ferrariCompilerDesignDistributed2021}, quantum entanglement resources\cite{meter2008arithmetic}.
However, comparing in the same environment is difficult because they evaluate performance in their respective architectures and constraints. 
Moreover, they did not distribute their results in reusable program formats like QIR (Quantum Intermediate Representation)\cite{mccaskeyMLIRDialectQuantum2021} because there is no programming language suitable for describing distributed quantum programs.

\subsection{Contribution}

This paper proposes InQuIR, an intermediate representation for interconnected quantum computers. Our key contributions are:
\begin{itemize}
\item We define the formal semantics of InQuIR, which has enough instructions to describe complicated behaviors of distributed quantum programs.
\item We give several examples written in InQuIR that helps us understand how distributed quantum programs work on quantum interconnects. These examples include runtime errors inherent in distributed systems, such as qubit memory exhaustion and deadlock in intercommunication between processors.
\item We provide a software tool using InQuIR to enable resource estimation for various distributed quantum programs.
  We also implement a toy compiler from traditional representation for quantum programs (e.g., OpenQASM) to InQuIR,
  but our tool does not depend on a particular compiler.
\item We give a roadmap for introducing static analysis that makes InQuIR programs more reliable. Specifically, we explain how to estimate resource consumption and address runtime errors such as deadlock using type systems.
\end{itemize}

The rest of this paper is organized as follows:
\Cref{sec:preliminaries} explains quantum interconnects, the background of this study.
\Cref{sec:related-work} gives the related work, including distributed programming languages and the libraries for distributed quantum computing.
\Cref{sec:design} discusses the language design of the InQuIR programming language.
\Cref{sec:inquir} introduces the formal syntax and semantics of the InQuIR programming language along with several examples.
\Cref{sec:compiler} provides our software tool to evaluate InQuIR programs.
\Cref{sec:verification} gives the roadmap on static verification of InQuIR.
\Cref{sec:discussion} discuss how to improve InQuIR for practical architectures and future work, and finally, \Cref{sec:conclusion} concludes this paper.

\section{Preliminaries}\label{sec:preliminaries}
\subsection{Quantum Interconnect}
Many state-of-the-art classical high-performance computing (HPC) systems \cite{frontier2021, dongarra2020report, lumi2022} utilize multiple processors and interconnects to achieve higher performance than a single processor. A similar approach may be practical for quantum computers. Such a quantum computing system can be realized by either providing a quantum communication channel called quantum interconnect \cite{Awschalom2021}. The quantum interconnect can transmit quantum states in a node to another node and/or distribute entanglement resources such as EPR state, GHZ state~\cite{greenberger1989going}, cluster state~\cite{raussendorf2001one}, and graph state~\cite{raussendorf2003measurement} between multiple quantum processors. This is expected to enable quantum computation using a large number of qubits. On the other hand, there is a concern that the quantum interconnect will become a bottleneck, leading to a slowdown in computational speed\cite{meter2008arithmetic}. To address this issue, several compilers \cite{zomorodi2018optimizing, daei2021improving, nikahd2021automated, dadkhah2021new, g2021efficient} have been studied to assign quantum circuits to distributed architecture so that the bottlenecks caused by quantum communication in the system are reduced.

In many cases, the data qubits used for quantum computation (matter qubits) and the communication qubits between participants (frying qubits) use qubits in different physical degrees of freedom. Devices for converting a qubit to a qubit of a different physical system are called quantum interfaces~\cite{hammerer2010quantum}, and their conversion efficiency and fidelity have improved remarkably in recent experiments~\cite{andrews2014bidirectional,mirhosseini2020superconducting,brubaker2022optomechanical,fan2018superconducting,sahu2022quantum,fernandez2019cavity,vogt2019efficient,kimble1998strong,girvin2009circuit}. It is expected that high-quality quantum interconnects will be implemented in the near future.

The quantum internet \cite{gisin2007quantum, kimble2008quantum} is also one of the methods to connect multiple quantum computers, but it differs from the quantum interconnect in several ways. The physical system of the quantum internet uses long-distance quantum communication, primarily using light, and thus must deal with the faults inherent in optical systems. This requires methods such as photonic quantum error correction codes\cite{gottesman2001encoding, michael2016new,grassl1999quantum} and quantum repeaters\cite{briegel1998quantum}. In addition, the cost of synchronization and classical communication between each node is greater than that of the interconnect, and complex multi-party communication protocols \cite{van2014quantum, wehner2018quantum, kozlowski2020designing} are required. Because of these difficulties, the quantum internet is expected to be more difficult to implement than the quantum interconnects.

\subsection{Teledata and Telegates}

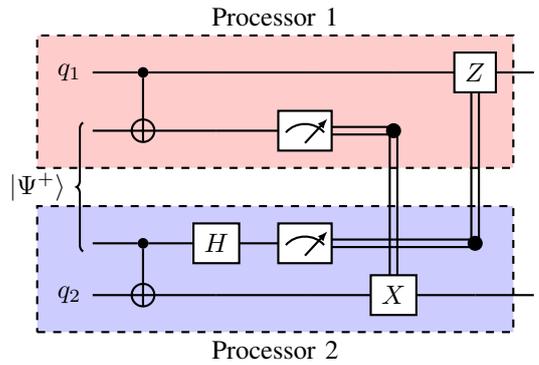
\begin{figure}[tb]
  \centering
  \begin{quantikz}
    \gategroup[wires=2,steps=7, style={dashed, inner sep=0.3em, fill=red!20}, background]{Processor 1} & \lstick{$q_1$} & \ctrl{1} & \qw      & \qw      & \qw       & \gate{Z}    & \qw \\[-1.0em]
    & \lstick[wires=2]{$\ket{\Psi^+}$} & \targ{}  & \qw      & \meter{} & \cwbend{2} &             & \\[1.0em]
    \gategroup[wires=2,steps=7, style={dashed, inner sep=0.3em, fill=blue!20}, label style={label position=below, anchor=north, yshift=-0.5em}, background]{Processor 2}&              & \ctrl{1} & \gate{H} & \meter{} & \cw       & \cwbend{-2} & \\[-1.0em]
    & \lstick{$q_2$} & \targ{}  & \qw      & \qw      & \gate{X}  & \qw         & \qw
  \end{quantikz}

  \caption{The remote $\mathit{CX}$ gate.}
  \label{fig:remote-cx}
\end{figure}

To run a quantum program on multiple processors,
remote operations between qubits placed on different processors are necessary.
We can use entanglements, which can be generated between directly connected quantum processors with quantum interconnects, to realize remote operations.
There are two ways to realize such operations by consuming entanglements: (1) \textit{teledata} transferring qubits by quantum teleportation~\cite{bennett1993teleporting},
and (2) \textit{telegates} achieved by gate-teleportation~\cite{gottesman1999teleport}.
In the first method, before applying a multi-qubit gate, one qubit is teleported to a processor where another qubit exists, and then the gate is applied locally.
In the second method, multi-qubit gates are applied to the target qubits remotely by gate teleportation; thus, the positions of qubits do not change.
For example, \Cref{fig:remote-cx} illustrates the remote $\mathit{CX}$ gate~\cite{jiang2007distributed} by gate-teleportation.
Distributed quantum compilers can use these strategies properly to generate an efficient quantum program.

\begin{figure*}[htbp]
  \begin{minipage}[htbp]{0.55\linewidth}
    \centering
    \includegraphics[width=7cm]{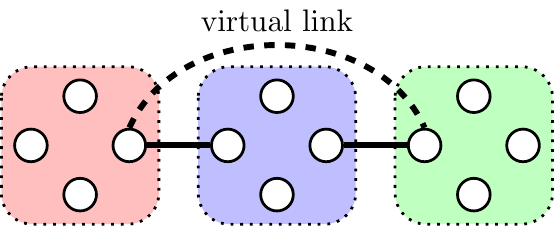}
  \end{minipage}
  \begin{minipage}[htbp]{0.4\linewidth}
    \begin{quantikz}
      \gategroup[wires=1, steps=5, style={dashed, inner xsep=0.7em, fill=red!25}, background]{}\lstick[wires=2]{$\ket{\Psi^+}$} & \qw      & \qw      & \gate{Z} \vcw{1}  & \qw \rstick[wires=4]{$\,\, \ket{\Psi^+}$} \\[0.8em]
      \gategroup[wires=2, steps=5, style={dashed, inner xsep=0.7em, fill=blue!25}, background]{} & \ctrl{1} & \gate{H} & \meter{}           & \\
      \lstick[wires=2]{$\ket{\Psi^+}$} & \targ{}    & \qw      & \meter{}            & \\[0.8em]
      \gategroup[wires=1, steps=5, style={dashed, inner xsep=0.7em, fill=green!25}, background]{}& \qw      & \qw      & \gate{X} \vcw{-1} & \qw
    \end{quantikz}
  \end{minipage}
  \caption{The topology of quantum processors (left) and the entanglement swapping procedure (right).}
  \label{fig:entanglement-swapping}
\end{figure*}

In general, the connection graph of quantum processors is not always complete. That is, some of them are not connected directly.
In this case, there is a procedure that creates an entanglement between distant processors by consuming multiple entanglements.
This procedure is called \textit{entanglement swapping}\cite{zukowski1993event}, as shown in \cref{fig:entanglement-swapping}.
In this example, there are three processors connected linearly, and the circuit generates an entanglement between endpoint processors by consuming two entanglements.
Finally, we can use the entanglement generated by entanglement swapping to implement a remote operation between distant quantum processors.

\section{Related Work}\label{sec:related-work}

\subsection{Programming Languages for Distributed Computing}
Dahlberg \etal presented NetQASM~\cite{dahlbergNetQASMLowlevelInstruction2021}, one of the variants of OpenQASM~\cite{cross2017open} that was designed for the quantum internet.
The NetQASM has a platform-independent instruction set, including entanglement generation and distribution of entanglements, and allows us to write a universal quantum protocol running on the quantum internet.
The difference between the NetQASM and the InQuIR is that the NetQASM is mainly designed for the quantum internet,
but the InQuIR is designed for quantum interconnects.
In addition, the InQuIR has formally defined semantics, while the NetQASM does not.
Formal semantics is important to discuss and reason about the behaviors of distributed quantum programs.

Gay and Nagarajan developed \emph{communicating quantum processes} (CQP)~\cite{gay2004CommunicatingQuantum} by extending $\pi$-calculus~\cite{milner1992CalculusMobile} with primitives for quantum information processing.
They also applied CQP to enable formal verification of quantum communication protocols, such as BB84~\cite{gay2005ProbabilisticModel,nagarajan2005AutomatedAnalysis}.
While CQP is suitable for describing quantum communication protocols in the quantum internet, it is not designed as an instruction set for controlling interconnected quantum processors.
We should develop a representation for quantum computer clusters and enable formal reasoning about distributed quantum algorithms.

The D$\pi$-calculus~\cite{hennessy2002ResourceAccessControl} is an extension of the $\pi$-calculus for distributed computing.
Unlike the $\pi$-calculus, the D$\pi$-calculus distinguishes between processes by the location where each process works.
Located processes are also useful to describe quantum programs on quantum interconnects
because each quantum processor may have different computational resources, such as the number of available qubits,
the connections with other processors, and the execution costs of instructions.
Thus, we designed the InQuIR as an extension of the D$\pi$-calculus with quantum primitives.

\subsection{The Frameworks for Distributed Quantum Computing}
H\"{a}ner \etal proposed the QMPI~\cite{hanerDistributedQuantumComputing2021} for implementing distributed quantum algorithms.
While we can widely use the QMPI to implement distributed quantum programs,
it is not suitable for use as a representation of the output destination of distributed quantum compilers
because the QMPI is just a framework, not a programming language.
They also presented the SENDQ model to evaluate the performance of distributed quantum algorithms.
Their model helps analyze distributed quantum programs, including InQuIR.

In recent years, various quantum Internet simulators\cite{dahlberg2018simulaqron, bartlett2018distributed, satoh2021quisp, coopmans2021netsquid,wu2021sequence,diadamo2021qunetsim} have been developed, which can validate communication protocols and simulate noises. Remarkably, NetQASM can be simulated on SimulaQron\cite{dahlberg2018simulaqron} and NetSquid\cite{coopmans2021netsquid}. These simulators could be useful in the design of the backend simulator of InQuIR.

\section{Language Design}\label{sec:design}
In considering intermediate representations on interconnected quantum computers, InQuIR was designed assuming the following model:

\noindent\textbf{Multiparty quantum interaction using EPR pairs.}
All quantum operations between multiple participants (telegate and teledata) are executed utilizing EPR pairs. For example, teleportation using EPR pairs is used to send and receive quantum states.
It is also possible to directly transmit and receive data qubits used in the calculation process without using any entangled state. However, this requires extremely high conversion efficiency and fidelity of the quantum interface, which is difficult to achieve, especially without quantum error-correcting codes. On the other hand, by using protocols such as entanglement purification, it is possible to achieve high-fidelity remote operations via entanglement resources.
In addition, the cost of program execution can be easily characterized by focusing on entanglement resource consumption.

\noindent\textbf{Architecture Independence.}
It is as independent of a particular physical system (e.g., superconductivity, trapped ions) or instruction set as possible. It does not deal with qubit allocation inside the participants, so it does not depend on constraints such as qubit connectivity. This enhances the reusability of distributed quantum programs written in InQuIR and also leaves room for relatively easy integration with the quantum internet simulators.

\noindent\textbf{Classical Operations and Communication.}
Distributed quantum programs running on different processors communicate classically with each other to execute their remote operations, including entanglement generations, teleportation, and remote CX gate.
Moreover, basic classical operations such as boolean operations are necessary to manage classical feedback of quantum measurements.
We design InQuIR as a language with primitives that allow us to express flexibly such classical operations and communication.
This makes it possible to use InQuIR to describe distributed measurement-based quantum computational models~\cite{danos2005DistributedMeasurementbasedQuantum}.

\noindent\textbf{Asynchronous and Concurrent Processes.}
In interconnected computers, a distributed program usually has several processes running concurrently on different processors. Thus InQuIR has to provide a primitive to express such concurrent processes.
Moreover, InQuIR adopts the asynchronous execution model; concurrent processes work asynchronously. Compared to the asynchronous model, the synchronous model may increase the execution time per instruction because it needs synchronization across all processors or a classical controller which manages the whole system.
In particular, in heterogeneous systems, where each processor adopts a different architecture or code space,
the slowest processor becomes the bottleneck of the entire system.
For this reason, the asynchronous execution model is appropriate to capture the behaviors of distributed systems. The asynchronous model also applies to the quantum internet if we use the InQuIR to write quantum communication protocols.

\section{The Syntax and Semantics of InQuIR}\label{sec:inquir}

\subsection{Syntax and Operational Semantics}

Firstly, we give the syntax of InQuIR defined by the grammar given in \cref{fig:syntax}.

\begin{figure*}[tb]
  \begin{gather*}
    \textit{Participants} \ni \inquirmv{p}
      \qquad\quad \textit{Sessions} \ni s
      \qquad\quad \textit{Labels} \ni l
      \qquad\quad \textit{DataQubits} \ni q
      \qquad\quad \textit{CommQubits} \ni \overline{q}
  \end{gather*}
  \begin{alignat*}{2}
    (\textit{Process})& &\quad \inquirmv{P} \Coloneqq\ &\inquirkw{stop} \pipe \inquirmv{s}  \inquirsym{=} \, \inquirkw{open} \, \inquirsym{[}  p_{{\mathrm{1}}}  \inquirsym{,} \, .. \, \inquirsym{,}  p_{\inquirmv{n}}  \inquirsym{]}  \inquirsym{;}  \inquirmv{P} \pipe \inquirkw{close} \, \inquirsym{(}  \inquirmv{s}  \inquirsym{)}  \inquirsym{;}  \inquirmv{P} \pipe \inquirmv{x}  \inquirsym{=} \, \inquirkw{init} \, \inquirsym{()}  \inquirsym{;}  \inquirmv{P} \pipe \inquirkw{free} \, \inquirmv{x}  \inquirsym{;}  \inquirmv{P} \pipe \\
    & & & \inquirmv{x}  \inquirsym{=}  \inquirnt{e}  \inquirsym{;}  \inquirmv{P} \pipe  U   \inquirsym{(}  \inquirnt{e_{{\mathrm{1}}}}  \inquirsym{,} \, .. \, \inquirsym{,}  \inquirnt{e_{\inquirmv{n}}}  \inquirsym{)}  \inquirsym{;}  \inquirmv{P} \pipe  \inquirmv{x}  = \mathbf{M}( \inquirnt{e_{{\mathrm{1}}}}  \inquirsym{,} \, .. \, \inquirsym{,}  \inquirnt{e_{\inquirmv{n}}} )   \inquirsym{;}  \inquirmv{P} \pipe \\
    & & & \inquirmv{x}  \inquirsym{=} \, \inquirkw{genEnt} \, \inquirsym{[}  \inquirmv{p}  \inquirsym{]}  \inquirsym{(}  \inquirmv{l}  \inquirsym{)}  \inquirsym{;}  \inquirmv{P} \pipe \inquirsym{(}  \inquirmv{x_{{\mathrm{1}}}}  \inquirsym{,}  \inquirmv{x_{{\mathrm{2}}}}  \inquirsym{)}  \inquirsym{=} \, \inquirkw{entSwap} \, \inquirsym{(}  \inquirnt{e_{{\mathrm{1}}}}  \inquirsym{,}  \inquirnt{e_{{\mathrm{2}}}}  \inquirsym{)}  \inquirsym{;}  \inquirmv{P} \pipe \\
    & & & \inquirkw{if} \, \inquirnt{e} \, \inquirkw{then} \, \inquirmv{P_{{\mathrm{1}}}} \, \inquirkw{else} \, \inquirmv{P_{{\mathrm{2}}}} \pipe \inquirmv{x_{{\mathrm{1}}}}  \inquirsym{=} \, \inquirkw{qrecv} \, \inquirsym{(}  \inquirmv{s}  \inquirsym{,}  \inquirmv{l}  \inquirsym{,}  \inquirnt{e_{{\mathrm{2}}}}  \inquirsym{)}  \inquirsym{;}  \inquirmv{P} \pipe \inquirkw{qsend} \, \inquirsym{[}  \inquirmv{p}  \inquirsym{]}  \inquirsym{(}  \inquirmv{s}  \inquirsym{,}  l  \inquirsym{,}  \inquirnt{e_{{\mathrm{1}}}}  \inquirsym{,}  \inquirnt{e_{{\mathrm{2}}}}  \inquirsym{)}  \inquirsym{;}  \inquirmv{P} \pipe \\
    & & & \inquirkw{rcxc} \, \inquirsym{[}  \inquirmv{p}  \inquirsym{]}  \inquirsym{(}  \inquirmv{s}  \inquirsym{,}  \inquirmv{l}  \inquirsym{,}  \inquirnt{e_{{\mathrm{1}}}}  \inquirsym{,}  \inquirnt{e_{{\mathrm{2}}}}  \inquirsym{)}  \inquirsym{;}  \inquirmv{P} \pipe \inquirkw{rcxt} \, \inquirsym{[}  \inquirmv{p}  \inquirsym{]}  \inquirsym{(}  \inquirmv{s}  \inquirsym{,}  \inquirmv{l}  \inquirsym{,}  \inquirnt{e_{{\mathrm{1}}}}  \inquirsym{,}  \inquirnt{e_{{\mathrm{2}}}}  \inquirsym{)}  \inquirsym{;}  \inquirmv{P} \pipe  \inquirmv{s} [  p  ]!( \inquirmv{l}  :  \inquirnt{e} )   \inquirsym{;}  \inquirmv{P} \pipe \inquirmv{s}  \inquirsym{\mbox{?}}  \inquirsym{(}  \inquirmv{l}  \inquirsym{:}  \inquirmv{x}  \inquirsym{)}  \inquirsym{;}  \inquirmv{P} \\
    (\textit{System})& & S \Coloneqq\ & \textbf{0}  \pipe  \llbracket  \inquirmv{P}  \rrbracket_{ \inquirmv{p} }  \pipe \inquirsym{(}  S_{{\mathrm{1}}}  \pipe  S_{{\mathrm{2}}}  \inquirsym{)} \\
    (\textit{Expression})& & \inquirnt{e} \Coloneqq\ &v \pipe  \inquirnt{e_{{\mathrm{1}}}}  \land  \inquirnt{e_{{\mathrm{2}}}}  \pipe  \inquirnt{e_{{\mathrm{1}}}}  \oplus  \inquirnt{e_{{\mathrm{2}}}}  \pipe \dots \\
    (\textit{Value})& & v \Coloneqq\ & 0  \pipe  1  \pipe q \pipe \overline{q} \pipe \inquirmv{x} \pipe \dots \\
    (\textit{Gate})& & U \Coloneqq\ &X \pipe Z \pipe H \pipe T \pipe \mathit{CX} \pipe \dots
  \end{alignat*}
  \caption{The syntax of the InQuIR.}
  \label{fig:syntax}
\end{figure*}

A number $\inquirmv{p}$ represents a participant (or processor).
A (classical) session $\inquirmv{s}$ is used for participants to send and receive their classical data between them.
A label $\inquirmv{l}$ identifies the kinds of communicated data and the partner of communication operations, as described later.
A session and a label can be considered to correspond to a communicator and a tag in the (Message Passing Interface) MPI library~\cite{10.5555/898758}, respectively.

In InQuIR, there are two kinds of qubits: \textit{communication qubits} and \textit{data qubits}.
A communication qubit plays a crucial role in quantum communication between processors, as explained in \cref{sec:preliminaries}.
Any qubits in all the other parts are data qubits.
$\textit{DataQubits}$ and $\textit{CommQubits}$ denote the set of data qubits and communication qubits, respectively.
We also write data qubits and communication qubits as $q$ and $\overline{q}$, respectively.

A process $\inquirmv{P}$ is a unit of an InQuIR program that has a sequence of operations.
A system $S$ represents an entire InQuIR program which consists of located processes working concurrently.
A located process $ \llbracket  \inquirmv{P}  \rrbracket_{ \inquirmv{p} } $ indicates that the process $\inquirmv{P}$ is executed on the participant $\inquirmv{p}$.
The composition of systems $S_{{\mathrm{1}}}  \pipe  S_{{\mathrm{2}}}$ represents that $S_{{\mathrm{1}}}$ and $S_{{\mathrm{2}}}$ are evaluated concurrently and asynchronously.
InQuIR allows $S_{{\mathrm{1}}}$ and $S_{{\mathrm{2}}}$ to contain processes which have the same location.
In other words, a concurrent system $ \llbracket  \inquirmv{P_{{\mathrm{1}}}}  \rrbracket_{ \inquirmv{p_{{\mathrm{1}}}} }   \pipe   \llbracket  \inquirmv{P_{{\mathrm{2}}}}  \rrbracket_{ \inquirmv{p_{{\mathrm{2}}}} } $ with $\inquirmv{p_{{\mathrm{1}}}} = \inquirmv{p_{{\mathrm{2}}}}$ is acceptable.
Note that the evaluation order of $\inquirmv{P_{{\mathrm{1}}}}$ and $\inquirmv{P_{{\mathrm{2}}}}$ in $ \llbracket  \inquirmv{P_{{\mathrm{1}}}}  \rrbracket_{ \inquirmv{p} }   \pipe   \llbracket  \inquirmv{P_{{\mathrm{2}}}}  \rrbracket_{ \inquirmv{p} } $ is non-deterministic.

Next, we introduce a \textit{runtime state} to define the operational semantics of InQuIR.

\begin{definition}{(\textit{Runtime state})}
  A runtime state of an InQuIR program is a $5$-tuple of $\rstate{\rho}{Q}{E}{\inquirmv{P}}{H}$, where
  \begin{itemize}
  \item $\rho$ is a density operator representing a quantum state,
  \item $Q$ is a set of available data qubits on each participant,
  \item $E$ is a set of available communication qubits between participants,
  \item $S$ is a system being evaluated currently, and
  \item $H$ is a heap for classical communication.
  \end{itemize}
\end{definition}
A heap $H : \textit{Sessions} \times \textit{Participants} \rightarrow \textit{List}\ (\textit{Labels} \times \textit{Values})$ manages classical data sent by other participants.
Each paticipants has a communication buffer $H(\inquirmv{s}, \inquirmv{p}) = \overline{\inquirmv{l}  \inquirsym{:}  \inquirmv{v}}$, where $\overline{\inquirmv{l}  \inquirsym{:}  \inquirmv{v}}$ is a sequence of labeled values. Labels represent for what purpose the value was sent.

Now we can define the operational semantics of InQuIR as a transition relation $\evalarrow$ on runtime states.
\begin{definition}{(\textit{Operational Semantics})}
  The operational semantics of InQuIR is defined by a transition relation $\evalarrow$ on runtime states.
  In other words, when we write $R \evalarrow R'$ for runtime states $R$ and $R'$, it can be read that the execution of $R$ transitions to $R'$ in one step.
  The details are given in \cref{fig:semantics}.
\end{definition}

\begin{figure*}[!h]
  \begin{gather*}
      (\textit{DataStore}) \quad Q \defeq \textit{Participants} \rightarrow 2^{\textit{DataQubits}}
      \qquad (\textit{EPRStore}) \quad E \defeq \textit{Participants} \rightharpoonup 2^{\textit{CommQubits}} \\
      (\textit{Heap})\quad H \defeq \textit{Session} \times \textit{Participant}
          \rightharpoonup \textit{List}\ (\textit{Labels} \times \textit{Values})
  \end{gather*}
  \begin{align*}
    &\rstate{\rho}{Q}{E}{ \llbracket  \inquirmv{s}  \inquirsym{=} \, \inquirkw{open} \, \inquirsym{[}  \inquirmv{p_{{\mathrm{1}}}}  \inquirsym{,} \, .. \, \inquirsym{,}  \inquirmv{p_{\inquirmv{n}}}  \inquirsym{]}  \inquirsym{;}  \inquirmv{P_{{\mathrm{1}}}}  \rrbracket_{ \inquirmv{p_{{\mathrm{1}}}} }   \pipe \, .. \, \pipe   \llbracket  \inquirmv{s}  \inquirsym{=} \, \inquirkw{open} \, \inquirsym{[}  \inquirmv{p_{{\mathrm{1}}}}  \inquirsym{,} \, .. \, \inquirsym{,}  \inquirmv{p_{\inquirmv{n}}}  \inquirsym{]}  \inquirsym{;}  \inquirmv{P_{\inquirmv{n}}}  \rrbracket_{ \inquirmv{p_{\inquirmv{n}}} } }{ H } \\
    &\qquad \evalarrow \rstate{\rho}{Q}{E}{ \llbracket  \inquirmv{P_{{\mathrm{1}}}}  \rrbracket_{ \inquirmv{p_{{\mathrm{1}}}} }   \pipe \, .. \, \pipe   \llbracket  \inquirmv{P_{\inquirmv{n}}}  \rrbracket_{ \inquirmv{p_{\inquirmv{n}}} } }{ H \uplus \inquirsym{\{}   ( \inquirmv{s} ,  \inquirmv{p_{{\mathrm{1}}}} ) \mapsto   \epsilon    \inquirsym{,} \, .. \, \inquirsym{,}   ( \inquirmv{s} ,  \inquirmv{p_{\inquirmv{n}}} ) \mapsto   \epsilon    \inquirsym{\}} }
  \end{align*}\vspace{-20pt}
  \begin{align*}
    \rstate{\rho}{Q}{E}{  \llbracket  \inquirkw{close} \, \inquirsym{(}  \inquirmv{s}  \inquirsym{)}  \inquirsym{;}  \inquirmv{P}  \rrbracket_{ \inquirmv{p} }  }{ H }
           \evalarrow \rstate{\rho}{Q}{E}{ \llbracket  \inquirmv{P}  \rrbracket_{ \inquirmv{p} } }{H \setminus \{ (\inquirmv{s}, \inquirmv{p}) \}} \\
  \end{align*}

  \vspace{-20pt}
  \infrule[]{
    \getq(Q, \inquirmv{p}) = (Q', q)
  }{
    \rstate{\rho}{Q}{E}{ \llbracket  \inquirmv{x}  \inquirsym{=} \, \inquirkw{init} \, \inquirsym{()}  \inquirsym{;}  \inquirmv{P}  \rrbracket_{ \inquirmv{p} } }{H}
           \evalarrow \rstate{\rho \otimes \ketbra{0}{q}{0}}{Q'}{E}{ \llbracket  \inquirsym{[}  q  \slash  \inquirmv{x}  \inquirsym{]} \, \inquirmv{P}  \rrbracket_{ \inquirmv{p} } }{H}
  }

  \vspace{-13pt}
  \begin{align*}
    \rstate{\rho}{Q}{E}{ \llbracket  \inquirkw{free} \, q  \inquirsym{;}  \inquirmv{P}  \rrbracket_{ \inquirmv{p} } }{H}
      \evalarrow& \rstate{\Tr_{q}(\rho)}{  \text{ret}( \inquirmv{Q} ,  \inquirmv{p} ,  q )  }{E}{ \llbracket  \inquirmv{P}  \rrbracket_{ \inquirmv{p} } }{H} \\
    \rstate{\rho}{Q}{E}{ \llbracket  \inquirkw{free} \, \overline{q}  \inquirsym{;}  \inquirmv{P}  \rrbracket_{ \inquirmv{p} } }{H}
    \evalarrow& \rstate{\Tr_{\overline{q}}(\rho)}{Q}{  \text{ret}( \inquirmv{E} ,  \inquirmv{p} ,  \overline{q} )  }{ \llbracket  \inquirmv{P}  \rrbracket_{ \inquirmv{p} } }{H}
  \end{align*}

  \infax[]{
    \rstate{\rho}{Q}{E}{ \llbracket  \inquirmv{x}  \inquirsym{=}  \inquirnt{e}  \inquirsym{;}  \inquirmv{P}  \rrbracket_{ \inquirmv{p} } }{H}
    \evalarrow \rstate{\rho}{Q}{E}{ \llbracket  \inquirsym{[}  \inquirmv{v}  \slash  \inquirmv{x}  \inquirsym{]} \, \inquirmv{P}  \rrbracket_{ \inquirmv{p} } }{H} \qquad (\text{if}~  \inquirnt{e}  \Downarrow  \inquirmv{v} )
  }

  \infrule[]{
    \forall \inquirmv{v_{\inquirmv{i}}} \in \textit{DataQubit} \cup \textit{CommQubit}
  }{
    \rstate{\rho}{Q}{E}{ \llbracket   U   \inquirsym{(}  \inquirmv{v_{{\mathrm{1}}}}  \inquirsym{,} \, .. \, \inquirsym{,}  \inquirmv{v_{\inquirmv{n}}}  \inquirsym{)}  \inquirsym{;}  \inquirmv{P}  \rrbracket_{ \inquirmv{p} } }{H}
           \evalarrow \rstate{U_{\inquirmv{v_{{\mathrm{1}}}}  \inquirsym{,} \, .. \, \inquirsym{,}  \inquirmv{v_{\inquirmv{n}}}}\rho U_{\inquirmv{v_{{\mathrm{1}}}}  \inquirsym{,} \, .. \, \inquirsym{,}  \inquirmv{v_{\inquirmv{n}}}}^\dagger}{Q}{E}{ \llbracket  \inquirmv{P}  \rrbracket_{ \inquirmv{p} } }{H}
  }
  \vspace{-8pt}
  \begin{gather*}
    \rstate{\rho}{Q}{E}{ \llbracket   \inquirmv{s} [  p_{{\mathrm{2}}}  ]!( \inquirmv{l}  :  \inquirmv{v} )   \inquirsym{;}  \inquirmv{P}  \rrbracket_{ \inquirmv{p_{{\mathrm{1}}}} } }{H}
           \evalarrow \rstate{\rho}{Q}{E}{  \llbracket  \inquirmv{P}  \rrbracket_{ \inquirmv{p_{{\mathrm{1}}}} }  }{  \text{push}( H ,  \inquirmv{s} ,  \inquirmv{p_{{\mathrm{2}}}} ,  \inquirmv{l}  :  \inquirmv{v} )  }
  \end{gather*}

  \infrule[]{
     \text{pop}( H ,  \inquirmv{s} ,  \inquirmv{p} ,  \inquirmv{l} )  = (H', \inquirmv{v})
  }{
    \rstate{\rho}{Q}{E}{  \llbracket  \inquirmv{s}  \inquirsym{\mbox{?}}  \inquirsym{(}  \inquirmv{l}  \inquirsym{:}  \inquirmv{x}  \inquirsym{)}  \inquirsym{;}  \inquirmv{P}  \rrbracket_{ \inquirmv{p} }  }{H}
    \evalarrow \rstate{\rho}{Q}{E}{ \llbracket  \inquirsym{[}  \inquirmv{v}  \slash  \inquirmv{x}  \inquirsym{]} \, \inquirmv{P}  \rrbracket_{ \inquirmv{p_{{\mathrm{1}}}} } }{ H' }
  }

  \infrule[]{
    \inquirmv{v} \in \{0, 1\}
    \andalso M_v = (I + (-1)^v Z_{q_{{\mathrm{1}}}}\cdots Z_{q_{\inquirmv{n}}})/2
  }{
    \rstate{\rho}{Q}{E}{ \llbracket   \inquirmv{x}  = \mathbf{M}( q_{{\mathrm{1}}}  \inquirsym{,} \, .. \, \inquirsym{,}  q_{\inquirmv{n}} )   \inquirsym{;}  \inquirmv{P}  \rrbracket_{ \inquirmv{p} } }{H}
    \evalarrow \rstate{M_v\rho M_v^\dagger}{Q}{E}{ \llbracket  \inquirsym{[}  \inquirmv{v}  \slash  \inquirmv{x}  \inquirsym{]} \, \inquirmv{P}  \rrbracket_{ \inquirmv{p} } }{H}
  }\vspace{-8pt}

  \infax[]{
    \rstate{\rho}{Q}{E}{ \llbracket  \inquirkw{if} \, \inquirnt{e} \, \inquirkw{then} \, \inquirmv{P_{{\mathrm{1}}}} \, \inquirkw{else} \, \inquirmv{P_{{\mathrm{2}}}}  \rrbracket_{ \inquirmv{p} } }{H}
           \evalarrow \rstate{\rho}{Q}{E}{ \llbracket  \inquirmv{P_{\inquirmv{i}}}  \rrbracket_{ \inquirmv{p} } }{H}
                      \qquad (\text{if}~  \inquirnt{e}  \Downarrow  \inquirmv{i}  \land \inquirmv{i} \in \{0, 1\})
  }

  \infrule[]{
     \text{get}( \inquirmv{E} ,  \inquirmv{p_{{\mathrm{1}}}} )  = (\inquirmv{E'}, \overline{q}_{{\mathrm{1}}})
    \andalso  \text{get}( \inquirmv{E'} ,  \inquirmv{p_{{\mathrm{2}}}} )  = (\inquirmv{E''}, \overline{q}_{{\mathrm{2}}})
  }{
    \rstate{\rho}{Q}{E}{ \llbracket  \inquirmv{x_{{\mathrm{1}}}}  \inquirsym{=} \, \inquirkw{genEnt} \, \inquirsym{[}  \inquirmv{p_{{\mathrm{2}}}}  \inquirsym{]}  \inquirsym{(}  \inquirmv{l}  \inquirsym{)}  \inquirsym{;}  \inquirmv{P_{{\mathrm{1}}}}  \rrbracket_{ \inquirmv{p_{{\mathrm{1}}}} }   \pipe   \llbracket  \inquirmv{x_{{\mathrm{2}}}}  \inquirsym{=} \, \inquirkw{genEnt} \, \inquirsym{[}  \inquirmv{p_{{\mathrm{1}}}}  \inquirsym{]}  \inquirsym{(}  \inquirmv{l}  \inquirsym{)}  \inquirsym{;}  \inquirmv{P_{{\mathrm{2}}}}  \rrbracket_{ \inquirmv{p_{{\mathrm{2}}}} } }{H} \\
    \evalarrow \rstate{\rho \otimes \ketbra{\Psi_+}{\overline{q}_{{\mathrm{1}}}\overline{q}_{{\mathrm{2}}}}{\Psi_+}}{Q}{E''}{ \llbracket  \inquirsym{[}  \overline{q}_{{\mathrm{1}}}  \slash  \inquirmv{x_{{\mathrm{1}}}}  \inquirsym{]} \, \inquirmv{P_{{\mathrm{1}}}}  \rrbracket_{ \inquirmv{p_{{\mathrm{1}}}} }   \pipe   \llbracket  \inquirsym{[}  \overline{q}_{{\mathrm{2}}}  \slash  \inquirmv{x_{{\mathrm{2}}}}  \inquirsym{]} \, \inquirmv{P_{{\mathrm{2}}}}  \rrbracket_{ \inquirmv{p_{{\mathrm{2}}}} } }{H}
  }

  \infax[]{
    \rstate{\rho}{Q}{E}{ \llbracket  \inquirsym{(}  \inquirmv{w_{{\mathrm{1}}}}  \inquirsym{,}  \inquirmv{w_{{\mathrm{2}}}}  \inquirsym{)}  \inquirsym{=} \, \inquirkw{entSwap} \, \inquirsym{(}  \overline{q}_{{\mathrm{1}}}  \inquirsym{,}  \overline{q}_{{\mathrm{2}}}  \inquirsym{)}  \inquirsym{;}  \inquirmv{P}  \rrbracket_{ \inquirmv{p} } }{H}
           \evalarrow \rstate{\rho}{Q}{ \text{ret}(  \text{ret}( \inquirmv{E} ,  \inquirmv{p} ,  \overline{q}_{{\mathrm{1}}} )  ,  \inquirmv{p} ,  \overline{q}_{{\mathrm{2}}} ) }{ \llbracket  \inquirmv{P}  \rrbracket_{ \inquirmv{p} } }{H\}\}}
  }

  \infrule[]{
    \inquirmv{Q'} =  \text{ret}( \inquirmv{Q} ,  \inquirmv{p_{{\mathrm{1}}}} ,  q ) 
    \andalso \inquirmv{E'} =  \text{ret}( \inquirmv{E} ,  \inquirmv{p_{{\mathrm{1}}}} ,  \overline{q} ) 
    \andalso H' = \hpush(H, \inquirmv{s}, \inquirmv{p_{{\mathrm{1}}}}, \inquirmv{p_{{\mathrm{2}}}}, \inquirmv{l}  \inquirsym{:}  \inquirmv{v_{{\mathrm{1}}}} \mdoubleplus \inquirmv{l}  \inquirsym{:}  \inquirmv{v_{{\mathrm{2}}}} ])
  }{
    \rstate{\rho}{Q}{E}{ \llbracket  \inquirkw{qsend} \, \inquirsym{[}  \inquirmv{p_{{\mathrm{2}}}}  \inquirsym{]}  \inquirsym{(}  \inquirmv{s}  \inquirsym{,}  \inquirmv{l}  \inquirsym{,}  q  \inquirsym{,}  \overline{q}  \inquirsym{)}  \inquirsym{;}  \inquirmv{P}  \rrbracket_{ \inquirmv{p_{{\mathrm{1}}}} } }{H}
      \evalarrow \rstate{\Tr_{q, \overline{q}}\left(U_{\mathrm{qsend}}\rho U_{\mathrm{qsend}}^\dagger\right)}{Q'}{E'}{ \llbracket  \inquirmv{P}  \rrbracket_{ \inquirmv{p_{{\mathrm{1}}}} } }{H'}
  }

  \infrule[]{
     \text{get}( \inquirmv{Q} ,  \inquirmv{p} )  = (\inquirmv{Q'}, q)
    \andalso \inquirmv{E'} =  \text{ret}( \inquirmv{E} ,  \inquirmv{p} ,  \overline{q} ) 
    \andalso  \text{pop}( H ,  \inquirmv{s} ,  \inquirmv{p} ,  \inquirmv{l} )  = (H', \inquirmv{v_{{\mathrm{1}}}})
    \andalso  \text{pop}( H' ,  \inquirmv{s} ,  \inquirmv{p} ,  \inquirmv{l} )  = (H'', \inquirmv{v_{{\mathrm{2}}}})
  }{
    \rstate{\rho}{Q}{E}{ \llbracket  \inquirmv{x}  \inquirsym{=} \, \inquirkw{qrecv} \, \inquirsym{(}  \inquirmv{s}  \inquirsym{,}  \inquirmv{l}  \inquirsym{,}  \overline{q}  \inquirsym{)}  \inquirsym{;}  \inquirmv{P}  \rrbracket_{ \inquirmv{p} } }{H}
    \evalarrow \rstate{U_{\mathrm{qrecv}}(\rho \otimes \ketbra{0}{q}{0})U_{\mathrm{qrecv}}^\dagger}{Q'}{E'}{ \llbracket  \inquirsym{[}  q  \slash  \inquirmv{x}  \inquirsym{]} \, \inquirmv{P}  \rrbracket_{ \inquirmv{p} } }{H''}
  }

  \infax[]{
    \rstate{\rho}{Q}{E}{ \llbracket  \inquirkw{rcxc} \, \inquirsym{[}  \inquirmv{p_{{\mathrm{2}}}}  \inquirsym{]}  \inquirsym{(}  \inquirmv{s}  \inquirsym{,}  \inquirmv{l}  \inquirsym{,}  q_{{\mathrm{1}}}  \inquirsym{,}  \overline{q}_{{\mathrm{1}}}  \inquirsym{)}  \inquirsym{;}  \inquirmv{P_{{\mathrm{1}}}}  \rrbracket_{ \inquirmv{p_{{\mathrm{1}}}} }   \pipe   \llbracket  \inquirkw{rcxt} \, \inquirsym{[}  \inquirmv{p_{{\mathrm{1}}}}  \inquirsym{]}  \inquirsym{(}  \inquirmv{s}  \inquirsym{,}  \inquirmv{l}  \inquirsym{,}  q_{{\mathrm{2}}}  \inquirsym{,}  \overline{q}_{{\mathrm{2}}}  \inquirsym{)}  \inquirsym{;}  \inquirmv{P_{{\mathrm{2}}}}  \rrbracket_{ \inquirmv{p_{{\mathrm{2}}}} } }{ H } \\
           \evalarrow \rstate{ \mathit{CX} _{\inquirmv{q_{{\mathrm{1}}}}\inquirmv{q_{{\mathrm{2}}}}}\rho \mathit{CX} _{\inquirmv{q_{{\mathrm{1}}}}\inquirmv{q_{{\mathrm{2}}}}}^\dagger}{Q}{ \text{ret}(  \text{ret}( \inquirmv{E} ,  \inquirmv{p_{{\mathrm{1}}}} ,  \overline{q}_{{\mathrm{1}}} )  ,  \inquirmv{p_{{\mathrm{2}}}} ,  \overline{q}_{{\mathrm{2}}} ) }{ \llbracket  \inquirmv{P_{{\mathrm{1}}}}  \rrbracket_{ \inquirmv{p_{{\mathrm{1}}}} }   \pipe   \llbracket  \inquirmv{P_{{\mathrm{2}}}}  \rrbracket_{ \inquirmv{p_{{\mathrm{2}}}} } }{H}
  }

  \begin{tabular}{cc}
    \begin{minipage}{0.40\hsize}
      \infrule[]{
        \rstate{\rho}{Q}{E}{S_{{\mathrm{1}}}}{H}
          \evalarrow \rstate{\rho'}{Q}{E}{S'_{{\mathrm{1}}}}{H}
      }{
        \rstate{\rho}{Q}{E}{S_{{\mathrm{1}}}  \pipe  S_{{\mathrm{2}}}}{H}
          \evalarrow \rstate{\rho'}{Q}{E}{S'_{{\mathrm{1}}}  \pipe  S_{{\mathrm{2}}}}{H}
      }
    \end{minipage}
    \begin{minipage}{0.60\hsize}
      \infrule[]{
        S_{{\mathrm{1}}} \equiv S_{{\mathrm{2}}}
        \andalso \rstate{\rho}{Q}{E}{S_{{\mathrm{1}}}}{H} \evalarrow \rstate{\rho'}{Q}{E}{S'_{{\mathrm{1}}}}{H'}
        \andalso S'_{{\mathrm{1}}} \equiv S'_{{\mathrm{2}}}
      }{
        \rstate{\rho}{Q}{E}{S_{{\mathrm{2}}}}{H} \evalarrow \rstate{\rho}{Q}{E}{S'_{{\mathrm{2}}}}{H}
      }
    \end{minipage}
  \end{tabular}

  \begin{tabular}{ccc}
    \begin{minipage}{0.25\hsize}
      \begin{gather*}
        S  \pipe   \textbf{0}  \equiv S
      \end{gather*}
    \end{minipage}
    \begin{minipage}{0.25\hsize}
      \begin{gather*}
        S_{{\mathrm{1}}}  \pipe  S_{{\mathrm{2}}} \equiv S_{{\mathrm{2}}}  \pipe  S_{{\mathrm{1}}}
      \end{gather*}
    \end{minipage}
    \begin{minipage}{0.25\hsize}
      \begin{gather*}
        S_{{\mathrm{1}}}  \pipe  \inquirsym{(}  S_{{\mathrm{2}}}  \pipe  S_{{\mathrm{3}}}  \inquirsym{)} \equiv \inquirsym{(}  S_{{\mathrm{1}}}  \pipe  S_{{\mathrm{2}}}  \inquirsym{)}  \pipe  S_{{\mathrm{3}}}
      \end{gather*}
    \end{minipage}
    \begin{minipage}{0.25\hsize}
      \begin{gather*}
         \llbracket  \inquirkw{stop}  \rrbracket_{ \inquirmv{p} }  \equiv  \textbf{0} 
      \end{gather*}
    \end{minipage}
  \end{tabular}

  \caption{The operational semantics of InQuIR.}
  \label{fig:semantics}
\end{figure*}

From here, we will give informal descriptions of the semantics of each operation in \cref{fig:syntax}.

\noindent\textbf{Stop.}
The stop operation $\inquirkw{stop}$ represents that a process is successfully finished.

\noindent\textbf{Open and Close.} The session opening operation $\inquirmv{s}  \inquirsym{=} \, \inquirkw{open} \, \inquirsym{[}  p_{{\mathrm{1}}}  \inquirsym{,} \, .. \, \inquirsym{,}  p_{\inquirmv{n}}  \inquirsym{]}$ opens a new session $\inquirmv{s}$ with processors $p_{{\mathrm{1}}}  \inquirsym{,} \, .. \, \inquirsym{,}  p_{\inquirmv{n}}$.
Processors use a session to send or receive their classical data, for example, the outcomes of quantum measurements.
When opening a new session, $n$ communication buffers are created in a current heap to store sending data through this session.
Conversely, the closing operation $\inquirkw{close} \, \inquirsym{(}  \inquirmv{s}  \inquirsym{)}$ closes a session $\inquirmv{s}$ and discards the buffer from the heap associated to the session $\inquirmv{s}$ and the participant $\inquirmv{p}$ which closes the session.

\noindent\textbf{Qubit Initialization.}
The qubit initialization $\inquirmv{x}  \inquirsym{=} \, \inquirkw{init} \, \inquirsym{()}$ obtains a qubit initialized with $\ket{0}$ from the current processor and binds a variable $\inquirmv{x}$ to it.
If a processor does not have an available qubit on qubit initialization, it waits for a qubit to be returned by a free operation explained in later.

Unlike quantum circuit representations, InQuIR does not specify the location of qubits on qubit initialization.
In other words, which qubits are allocated on each qubit initialization is left to a more architecture-specific backend compiler or the runtime scheduler.

\begin{figure}[tb]
  \centering
  \includegraphics[width=7cm]{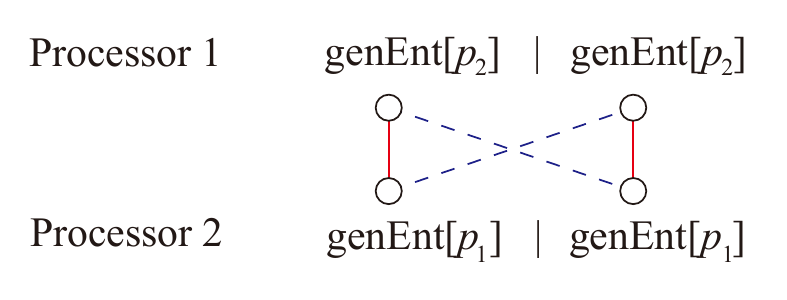}
  \caption{Ambigious entanglement generations.}
  \label{fig:ambiguous-genEnt}
\end{figure}

\noindent\textbf{Entanglement Generation.} The EPR pair generation $\inquirmv{x}  \inquirsym{=} \, \inquirkw{genEnt} \, \inquirsym{[}  \inquirmv{p}  \inquirsym{]}  \inquirsym{(}  \inquirmv{l}  \inquirsym{)}$ generates an EPR pair with anothor participant $\inquirmv{p}$ and binds a variable $\inquirmv{x}$ to it. The label $\inquirmv{l}$ identifies its partner operation running on the participant $\inquirmv{p}$ to avoid ambiguity as shown in \cref{fig:ambiguous-genEnt}. 
If two pairs of entanglement generations are running concurrently on each processor,
there are two possible connections between them (the red lines and blue dotted lines).

\noindent\textbf{Qubit Free.}
The free operation $\inquirkw{free} \, \inquirmv{x}$ returns a data qubit or communication qubit $\inquirmv{x}$ to the processor.
This operation takes the partial trace of the current state with respect to the qubit $\inquirmv{x}$.
At this time, InQuIR believes that the qubit $\inquirmv{x}$ can be deallocated safely; that is, the qubit $\inquirmv{x}$ is disentangled with any other qubits.
In general, it is difficult to check whether a given qubit can be reset safely~\cite{bichsel2020SilqHighlevelQuantum}.
The guarantee of safe resetting is out of the scope of InQuIR.

\noindent\textbf{Entanglement Swapping.}
The entanglement swapping expression $\inquirsym{(}  \inquirmv{x_{{\mathrm{1}}}}  \inquirsym{,}  \inquirmv{x_{{\mathrm{2}}}}  \inquirsym{)}  \inquirsym{=} \, \inquirkw{entSwap} \, \inquirsym{(}  \inquirmv{y_{{\mathrm{1}}}}  \inquirsym{,}  \inquirmv{y_{{\mathrm{2}}}}  \inquirsym{)}$ uses two entanglements $\inquirmv{y_{{\mathrm{1}}}}$ and $\inquirmv{y_{{\mathrm{2}}}}$ to create a virtual link by the entanglement swapping. This operation binds the outcomes of the measurements in the entanglement swapping procedure to the variables $\inquirmv{x_{{\mathrm{1}}}}$ and $\inquirmv{x_{{\mathrm{2}}}}$.
\cref{fig:entswap-stmt} illustrates how $\inquirsym{(}  \inquirmv{x_{{\mathrm{1}}}}  \inquirsym{,}  \inquirmv{x_{{\mathrm{2}}}}  \inquirsym{)}  \inquirsym{=} \, \inquirkw{entSwap} \, \inquirsym{(}  \inquirmv{y_{{\mathrm{1}}}}  \inquirsym{,}  \inquirmv{y_{{\mathrm{2}}}}  \inquirsym{)}$ works.
In order to complete the whole entanglement swapping, we have to ensure that the endpoints receive $\inquirmv{x_{{\mathrm{1}}}}$ (or $\inquirmv{x_{{\mathrm{2}}}}$) to apply a controlled $Z$ (or $X$) gate.

\begin{figure}[htb]
  \centering
  \begin{quantikz}
     \lstick{$y_1$} & \ctrl{1} & \gate{H} & \meter{$x_1$} \\[-0.4em]
     \lstick{$y_2$} & \targ{}  & \qw      & \meter{$x_2$}
  \end{quantikz}

  \caption{Illustration of $\inquirsym{(}  \inquirmv{x_{{\mathrm{1}}}}  \inquirsym{,}  \inquirmv{x_{{\mathrm{2}}}}  \inquirsym{)}  \inquirsym{=} \, \inquirkw{entSwap} \, \inquirsym{(}  \inquirmv{y_{{\mathrm{1}}}}  \inquirsym{,}  \inquirmv{y_{{\mathrm{2}}}}  \inquirsym{)}$.}
  \label{fig:entswap-stmt}
\end{figure}

\noindent\textbf{Measurements.}
The operation $ \inquirmv{x}  = \mathbf{M}( \inquirmv{x_{{\mathrm{1}}}}  \inquirsym{,} \, .. \, \inquirsym{,}  \inquirmv{x_{\inquirmv{n}}} ) $ performs a quantum measurement $\{M_s\}$ and binds $\inquirmv{x}$ to the classical outcome, where
\begin{gather*}
  M_s = \frac{I + (-1)^s Z_{x_1} \dots Z_{x_n}}{2}\quad (s = 0, 1)
\end{gather*}

\noindent\textbf{Qubit Teleportation.}
The qubit sending operation $\inquirkw{qsend} \, \inquirsym{[}  \inquirmv{p}  \inquirsym{]}  \inquirsym{(}  \inquirmv{s}  \inquirsym{,}  \inquirmv{l}  \inquirsym{,}  \inquirmv{x_{{\mathrm{1}}}}  \inquirsym{,}  \inquirmv{x_{{\mathrm{2}}}}  \inquirsym{)}$ teleports a data qubit $\inquirmv{x_{{\mathrm{1}}}}$ through an entanglement $\inquirmv{x_{{\mathrm{2}}}}$. Similarly, the qubit receiving operation $\inquirmv{x_{{\mathrm{1}}}}  \inquirsym{=} \, \inquirkw{qrecv} \, \inquirsym{(}  \inquirmv{s}  \inquirsym{,}  \inquirmv{l}  \inquirsym{,}  \inquirmv{x_{{\mathrm{2}}}}  \inquirsym{)}$ receives a data qubit througn an entanglement $\inquirmv{x_{{\mathrm{2}}}}$ and binds a variable $\inquirmv{x_{{\mathrm{1}}}}$ to it. The session $\inquirmv{s}$ and the label $\inquirmv{l}$ are used in the classical communication for Pauli correction in qubit teleportation. These operations can be written as follows:
\begin{align*}
  &\inquirkw{qsend} \, \inquirsym{[}  \inquirmv{p_{{\mathrm{2}}}}  \inquirsym{]}  \inquirsym{(}  \inquirmv{s}  \inquirsym{,}  \inquirmv{l}  \inquirsym{,}  \inquirmv{x_{{\mathrm{1}}}}  \inquirsym{,}  \inquirmv{x_{{\mathrm{2}}}}  \inquirsym{)} \\
  &\quad\defeq  \mathit{CX}   \inquirsym{(}  \inquirmv{x_{{\mathrm{1}}}}  \inquirsym{,}  \inquirmv{x_{{\mathrm{2}}}}  \inquirsym{)}  \inquirsym{;}   H   \inquirsym{(}  \inquirmv{x_{{\mathrm{1}}}}  \inquirsym{)}  \inquirsym{;}   \inquirmv{y_{{\mathrm{1}}}}  = \mathbf{M}( \inquirmv{x_{{\mathrm{1}}}} )   \inquirsym{;}   \inquirmv{y_{{\mathrm{2}}}}  = \mathbf{M}( \inquirmv{x_{{\mathrm{2}}}} ) ; \\
  &\quad\qquad  \inquirmv{s} [  \inquirmv{p_{{\mathrm{2}}}}  ]!( \inquirmv{l}  :  \inquirmv{y_{{\mathrm{1}}}} )   \inquirsym{;}   \inquirmv{s} [  \inquirmv{p_{{\mathrm{2}}}}  ]!( \inquirmv{l}  :  \inquirmv{y_{{\mathrm{2}}}} )   \inquirsym{;}  \inquirkw{free} \, \inquirmv{x_{{\mathrm{1}}}}  \inquirsym{;}  \inquirkw{free} \, \inquirmv{x_{{\mathrm{2}}}} \\
  &\inquirmv{x_{{\mathrm{1}}}}  \inquirsym{=} \, \inquirkw{qrecv} \, \inquirsym{(}  \inquirmv{s}  \inquirsym{,}  \inquirmv{l}  \inquirsym{,}  \inquirmv{x_{{\mathrm{2}}}}  \inquirsym{)} \\
  &\quad\defeq \inquirmv{x_{{\mathrm{1}}}}  \inquirsym{=} \, \inquirkw{init} \, \inquirsym{()}  \inquirsym{;}  \inquirmv{s}  \inquirsym{\mbox{?}}  \inquirsym{(}  \inquirmv{l}  \inquirsym{:}  \inquirmv{y_{{\mathrm{1}}}}  \inquirsym{)}  \inquirsym{;}  \inquirmv{s}  \inquirsym{\mbox{?}}  \inquirsym{(}  \inquirmv{l}  \inquirsym{:}  \inquirmv{y_{{\mathrm{2}}}}  \inquirsym{)}; \\
  &\quad\qquad Z^{y_1}(x_2); X^{y_2}(x_2); \mathit{SWAP}(\inquirmv{x_{{\mathrm{1}}}}  \inquirsym{,}  \inquirmv{x_{{\mathrm{2}}}}); \inquirkw{free} \, \inquirmv{x_{{\mathrm{2}}}}
\end{align*}
Note that $U^v$ denotes a unitary operation controlled by a classical value $\inquirmv{v}$.
The qubit sending operation does not need to wait for a receiver because the quantum teleportation is one-way communication.

\noindent\textbf{Remote CX Gate.}
The remote CX operations $\inquirkw{rcxc} \, \inquirsym{[}  \inquirmv{p}  \inquirsym{]}  \inquirsym{(}  \inquirmv{s}  \inquirsym{,}  \inquirmv{l}  \inquirsym{,}  \inquirmv{x_{{\mathrm{1}}}}  \inquirsym{,}  \inquirmv{x_{{\mathrm{2}}}}  \inquirsym{)}$ and $\inquirkw{rcxt} \, \inquirsym{[}  \inquirmv{p}  \inquirsym{]}  \inquirsym{(}  \inquirmv{s}  \inquirsym{,}  \inquirmv{l}  \inquirsym{,}  \inquirmv{x_{{\mathrm{3}}}}  \inquirsym{,}  \inquirmv{x_{{\mathrm{4}}}}  \inquirsym{)}$ apply the CX gate remotely to $\inquirmv{x_{{\mathrm{1}}}}$ (the controlled qubit) and $\inquirmv{x_{{\mathrm{3}}}}$ (the target qubit) by consuming the entanglement pair $(\inquirmv{x_{{\mathrm{2}}}}, \inquirmv{x_{{\mathrm{4}}}})$.
These two instructions can also be written as follows:
\begin{align*}
  &\inquirkw{rcxc} \, \inquirsym{[}  \inquirmv{p}  \inquirsym{]}  \inquirsym{(}  \inquirmv{s}  \inquirsym{,}  \inquirmv{l}  \inquirsym{,}  \inquirmv{x_{{\mathrm{1}}}}  \inquirsym{,}  \inquirmv{x_{{\mathrm{2}}}}  \inquirsym{)} \defeq  \mathit{CX}   \inquirsym{(}  \inquirmv{x_{{\mathrm{1}}}}  \inquirsym{,}  \inquirmv{x_{{\mathrm{2}}}}  \inquirsym{)};  \inquirmv{y}  = \mathbf{M}( \inquirmv{x_{{\mathrm{2}}}} ) ; \\
  &\qquad\qquad \inquirmv{s} [  \inquirmv{p}  ]!( \inquirmv{l}  :  \inquirmv{y} ) ; \inquirmv{s}  \inquirsym{\mbox{?}}  \inquirsym{(}  \inquirmv{l}  \inquirsym{:}  \inquirmv{y'}  \inquirsym{)}; Z^{y'}(\inquirmv{x_{{\mathrm{1}}}}) \\
  &\inquirkw{rcxt} \, \inquirsym{[}  \inquirmv{p}  \inquirsym{]}  \inquirsym{(}  \inquirmv{s}  \inquirsym{,}  \inquirmv{l}  \inquirsym{,}  \inquirmv{x_{{\mathrm{1}}}}  \inquirsym{,}  \inquirmv{x_{{\mathrm{2}}}}  \inquirsym{)} \defeq  \mathit{CX}   \inquirsym{(}  \inquirmv{x_{{\mathrm{2}}}}  \inquirsym{,}  \inquirmv{x_{{\mathrm{1}}}}  \inquirsym{)};  H   \inquirsym{(}  \inquirmv{x_{{\mathrm{2}}}}  \inquirsym{)};  \inquirmv{y'}  = \mathbf{M}( \inquirmv{x_{{\mathrm{2}}}} ) ; \\
  &\qquad\qquad \inquirmv{s} [  \inquirmv{p_{{\mathrm{1}}}}  ]!( \inquirmv{l}  :  \inquirmv{y'} ) ; \inquirmv{s}  \inquirsym{\mbox{?}}  \inquirsym{(}  \inquirmv{l}  \inquirsym{:}  \inquirmv{y}  \inquirsym{)}; X^{y}(\inquirmv{x_{{\mathrm{1}}}}) \\
\end{align*}

\noindent\textbf{Classical Communication.}
The sending operation $ \inquirmv{s} [  \inquirmv{p}  ]!( \inquirmv{l}  :  \inquirmv{v} ) $ sends a classical data $\inquirmv{v}$ labeled with $\inquirmv{l}$ to a processor $p$ through a session $\inquirmv{s}$. The sending data is appended to the end of the communication buffer $H(s, \inquirmv{p})$.
The receiving operation $\inquirmv{s}  \inquirsym{\mbox{?}}  \inquirsym{(}  \inquirmv{l}  \inquirsym{:}  \inquirmv{x}  \inquirsym{)}$ obtains a data labeled with $\inquirmv{l}$ from the communication buffer and binds $\inquirmv{x}$ to it. If there are two or more data labeled with $\inquirmv{l}$ in the buffer, the first one is popped. If the target data has not been sent yet, this operation waits for it.

\noindent\textbf{Branch.}
The branching operation $\inquirkw{if} \, \inquirnt{e} \, \inquirkw{then} \, \inquirmv{P_{{\mathrm{1}}}} \, \inquirkw{else} \, \inquirmv{P_{{\mathrm{2}}}}$ evaluates $\inquirnt{e}$ and choose $\inquirmv{P_{{\mathrm{1}}}}$ or $\inquirmv{P_{{\mathrm{2}}}}$ according to the result of the evaluation. The notation $ \inquirnt{e}  \Downarrow  \inquirmv{v} $ indicates that $\inquirnt{e}$ is evaluated to a value $\inquirmv{v}$.

Next, we introduce notations used in the operational semantics.

$\inquirsym{[}  \inquirmv{v}  \slash  \inquirmv{x}  \inquirsym{]}$ denotes the substitution of $\inquirmv{v}$ for $\inquirmv{x}$ in processes and expressions.
Formally, the substitution is defined as follows:
\begin{align*}
  \inquirsym{[}  \inquirmv{v}  \slash  \inquirmv{x}  \inquirsym{]}  \inquirmv{y} &\defeq \begin{cases}
    y & (x \neq y) \\
    v & (x = y)
  \end{cases} \\
  \inquirsym{[}  \inquirmv{v}  \slash  \inquirmv{x}  \inquirsym{]} \, \inquirmv{y}  \inquirsym{=}  \inquirnt{e}  \inquirsym{;}  \inquirmv{P} &\defeq \begin{cases}
    \inquirmv{y}  \inquirsym{=}  \inquirsym{[}  \inquirmv{v}  \slash  \inquirmv{x}  \inquirsym{]}  \inquirnt{e}  \inquirsym{;}  \inquirsym{[}  \inquirmv{v}  \slash  \inquirmv{x}  \inquirsym{]} \, \inquirmv{P} & (x \neq y) \\
    \inquirmv{y}  \inquirsym{=}  \inquirsym{[}  \inquirmv{v}  \slash  \inquirmv{x}  \inquirsym{]}  \inquirnt{e}  \inquirsym{;}  \inquirmv{P} & (x = y)
  \end{cases} \\
  &\vdots
\end{align*}

We write $ \text{get}( \inquirmv{Q} ,  \inquirmv{p} )  = (Q', q)$ if and only if all the following conditions hold:
\begin{itemize}
\item $q \in Q(\inquirmv{p})$,
\item $Q'(\inquirmv{p}) = Q(\inquirmv{p}) \setminus \{q\}$, and
\item $\forall \inquirmv{p'}.\ \inquirmv{p} \neq \inquirmv{p'} \Rightarrow Q(\inquirmv{p'}) = Q'(\inquirmv{p'})$.
\end{itemize}
We write $Q' =  \text{ret}( \inquirmv{Q} ,  \inquirmv{p} ,  q ) $ if and only if all the following conditions hold:
\begin{itemize}
\item $Q'(\inquirmv{p}) = Q(\inquirmv{p}) \cup \{q\}$, and
\item $\forall \inquirmv{p'}.\ \inquirmv{p} \neq \inquirmv{p'} \Rightarrow Q(\inquirmv{p'}) = Q'(\inquirmv{p'})$.
\end{itemize}
We also use $\text{get}$ and $\text{ret}$ for communication qubits $E$ in the same manner.
We write $H' =  \text{push}( H ,  \inquirmv{s} ,  \inquirmv{p} ,  \inquirmv{l}  :  \inquirmv{v} ) $ if and only if all the following conditions hold:
\begin{itemize}
  \item $H'(\inquirmv{s}, \inquirmv{p}) = H(\inquirmv{s}, \inquirmv{p}) \mdoubleplus \inquirmv{l}  \inquirsym{:}  \inquirmv{v}$, and
\item $\forall \inquirmv{s'}, \inquirmv{p'}.\ \inquirmv{s} \neq \inquirmv{s'} \lor \inquirmv{p} \neq \inquirmv{p'} \Rightarrow H(s, p) = H(\inquirmv{s'}, \inquirmv{p'})$,
\end{itemize}
where the operator $\mdoubleplus$ concatenates two lists.
We write $ \text{pop}( H ,  \inquirmv{s} ,  \inquirmv{p} ,  \inquirmv{l} )  = (H', \inquirmv{v})$ if and only if all the following conditions hold:
\begin{itemize}
\item $H(s, p) = \overline{\inquirmv{l_{{\mathrm{1}}}}  \inquirsym{:}  \inquirmv{v_{{\mathrm{1}}}}} \mdoubleplus \inquirmv{l}  \inquirsym{:}  \inquirmv{v} \mdoubleplus \overline{\inquirmv{l_{{\mathrm{2}}}}  \inquirsym{:}  \inquirmv{v_{{\mathrm{2}}}}}$
  and there does not exist $\inquirmv{v'}$ such that $\inquirmv{l}  \inquirsym{:}  \inquirmv{v'} \in \overline{\inquirmv{l_{{\mathrm{1}}}}  \inquirsym{:}  \inquirmv{v_{{\mathrm{1}}}}}$,
\item $H'(s, p) = \overline{\inquirmv{l_{{\mathrm{1}}}}  \inquirsym{:}  \inquirmv{v_{{\mathrm{1}}}}} \mdoubleplus \overline{\inquirmv{l_{{\mathrm{2}}}}  \inquirsym{:}  \inquirmv{v_{{\mathrm{2}}}}}$, and
\item $\forall \inquirmv{s'}, \inquirmv{p'}.\ \inquirmv{s} \neq \inquirmv{s'} \lor \inquirmv{p} \neq \inquirmv{p'} \Rightarrow H(s, p) = H(\inquirmv{s'}, \inquirmv{p'})$.
\end{itemize}

Finally, we define a \textit{stuck state} which plays a crucial role in the safety of InQuIR programs.
\begin{definition}{(\textit{Stuck state)}}
  If $\inquirmv{P} \not\equiv  \textbf{0} $
  and there does not exists a runtime state $\rstate{\rho'}{Q'}{E'}{\inquirmv{P'}}{H'}$ such that
  $\rstate{\rho}{Q}{E}{\inquirmv{P}}{H} \evalarrow \rstate{\rho'}{Q'}{E'}{\inquirmv{P'}}{H'}$,
  then the runtime state $\rstate{\rho}{Q}{E}{\inquirmv{P}}{H}$ called stuck state.
\end{definition}
A stuck state is undesirable because it fails to finish all processes.
A distributed program can reach this state due to the limited computational resources, as explained later.
Thus, distributed quantum compilers have to transpile a quantum program so that the output does not get stuck.
The notion of a stuck state enables us to discuss the safety of distributed quantum programs.
That is, if a distributed quantum program must not reach a stuck state, it can be executed safely.
This notion is quite important in the static verification of InQuIR,
which checks whether we can run an InQuIR program safely or not.

\subsection{Examples}

\begin{example}{(Entanglement Swapping)}
  The first example shows how to apply the remote CX gate between distant processors by entanglement swapping.
  \begin{align*}
    E &\defeq \{(\inquirmv{p_{{\mathrm{1}}}}, \inquirmv{p_{{\mathrm{2}}}}) \mapsto \{\overline{q}_{{\mathrm{1}}}\}, (\inquirmv{p_{{\mathrm{2}}}}, \inquirmv{p_{{\mathrm{1}}}}) \mapsto \{\overline{q}_{{\mathrm{2}}}\}, \\
      &\qquad\qquad (\inquirmv{p_{{\mathrm{2}}}}, \inquirmv{p_{{\mathrm{3}}}}) \mapsto \{\overline{q}_{{\mathrm{3}}}\}, (\inquirmv{p_{{\mathrm{3}}}}, \inquirmv{p_{{\mathrm{2}}}}) \mapsto \{\overline{q}_{{\mathrm{4}}}\} \} \\
    \inquirmv{P_{{\mathrm{1}}}} &\defeq \inquirmv{s}  \inquirsym{=} \, \inquirkw{open} \, \inquirsym{[}  \inquirmv{p_{{\mathrm{1}}}}  \inquirsym{,}  \inquirmv{p_{{\mathrm{2}}}}  \inquirsym{,}  \inquirmv{p_{{\mathrm{3}}}}  \inquirsym{]}; \inquirmv{x_{{\mathrm{2}}}}  \inquirsym{=} \, \inquirkw{genEnt} \, \inquirsym{[}  \inquirmv{p_{{\mathrm{2}}}}  \inquirsym{]}  \inquirsym{(}  \inquirmv{l_{{\mathrm{1}}}}  \inquirsym{)}; \\
      &\qquad \inquirmv{s}  \inquirsym{\mbox{?}}  \inquirsym{(}  \inquirmv{l}  \inquirsym{:}  \inquirmv{w_{{\mathrm{1}}}}  \inquirsym{)};Z^{w_1}(\inquirmv{x_{{\mathrm{2}}}}); \inquirkw{rcxc} \, \inquirsym{[}  \inquirmv{p_{{\mathrm{3}}}}  \inquirsym{]}  \inquirsym{(}  \inquirmv{s}  \inquirsym{,}  \inquirmv{l_{{\mathrm{3}}}}  \inquirsym{,}  q_{{\mathrm{1}}}  \inquirsym{,}  \inquirmv{x_{{\mathrm{2}}}}  \inquirsym{)} \\
    \inquirmv{P_{{\mathrm{2}}}} &\defeq \inquirmv{s}  \inquirsym{=} \, \inquirkw{open} \, \inquirsym{[}  \inquirmv{p_{{\mathrm{1}}}}  \inquirsym{,}  \inquirmv{p_{{\mathrm{2}}}}  \inquirsym{,}  \inquirmv{p_{{\mathrm{3}}}}  \inquirsym{]}; \inquirmv{z_{{\mathrm{1}}}}  \inquirsym{=} \, \inquirkw{genEnt} \, \inquirsym{[}  \inquirmv{p_{{\mathrm{1}}}}  \inquirsym{]}  \inquirsym{(}  \inquirmv{l_{{\mathrm{1}}}}  \inquirsym{)}; \\
    &\qquad  \inquirmv{z_{{\mathrm{2}}}}  \inquirsym{=} \, \inquirkw{genEnt} \, \inquirsym{[}  \inquirmv{p_{{\mathrm{3}}}}  \inquirsym{]}  \inquirsym{(}  \inquirmv{l_{{\mathrm{2}}}}  \inquirsym{)}; \\
    &\qquad \inquirsym{(}  \inquirmv{w_{{\mathrm{1}}}}  \inquirsym{,}  \inquirmv{w_{{\mathrm{2}}}}  \inquirsym{)}  \inquirsym{=} \, \inquirkw{entSwap} \, \inquirsym{(}  \inquirmv{z_{{\mathrm{1}}}}  \inquirsym{,}  \inquirmv{z_{{\mathrm{2}}}}  \inquirsym{)}; \\
           &\qquad  \inquirmv{s} [  \inquirmv{p_{{\mathrm{1}}}}  ]!( \inquirmv{l}  :  \inquirmv{w_{{\mathrm{1}}}} )   \inquirsym{;}   \inquirmv{s} [  \inquirmv{p_{{\mathrm{2}}}}  ]!( \inquirmv{l}  :  \inquirmv{w_{{\mathrm{2}}}} )  \\
    \inquirmv{P_{{\mathrm{3}}}} &\defeq \inquirmv{s}  \inquirsym{=} \, \inquirkw{open} \, \inquirsym{[}  \inquirmv{p_{{\mathrm{1}}}}  \inquirsym{,}  \inquirmv{p_{{\mathrm{2}}}}  \inquirsym{,}  \inquirmv{p_{{\mathrm{3}}}}  \inquirsym{]}; \inquirmv{y_{{\mathrm{2}}}}  \inquirsym{=} \, \inquirkw{genEnt} \, \inquirsym{[}  \inquirmv{p_{{\mathrm{2}}}}  \inquirsym{]}  \inquirsym{(}  \inquirmv{l_{{\mathrm{2}}}}  \inquirsym{)}; \\
      &\qquad \inquirmv{s}  \inquirsym{\mbox{?}}  \inquirsym{(}  \inquirmv{l}  \inquirsym{:}  \inquirmv{w_{{\mathrm{2}}}}  \inquirsym{)}; X^{w_2}(\inquirmv{y_{{\mathrm{2}}}}); \inquirkw{rcxt} \, \inquirsym{[}  \inquirmv{p_{{\mathrm{1}}}}  \inquirsym{]}  \inquirsym{(}  \inquirmv{s}  \inquirsym{,}  \inquirmv{l_{{\mathrm{3}}}}  \inquirsym{,}  q_{{\mathrm{2}}}  \inquirsym{,}  \inquirmv{y_{{\mathrm{2}}}}  \inquirsym{)} \\
    S &\defeq   \llbracket  \inquirmv{P_{{\mathrm{1}}}}  \rrbracket_{ \inquirmv{p_{{\mathrm{1}}}} }   \pipe   \llbracket  \inquirmv{P_{{\mathrm{2}}}}  \rrbracket_{ \inquirmv{p_{{\mathrm{2}}}} }   \,|\,   \llbracket  \inquirmv{P_{{\mathrm{3}}}}  \rrbracket_{ \inquirmv{p_{{\mathrm{3}}}} }  
  \end{align*}
\end{example}
Note that the final operation $\inquirkw{stop}$ is omitted here for simplicity, and the same applies in all examples provided later.
In this example, we assume that qubits $q_{{\mathrm{1}}}  \inquirsym{,}  q_{{\mathrm{2}}}$ have already been initialized.
In addition, processors $\inquirmv{p_{{\mathrm{1}}}}, \inquirmv{p_{{\mathrm{2}}}}, \inquirmv{p_{{\mathrm{3}}}}$ are connected linearly, and exactly one communication qubit is available at each connection.
\Cref{tbl:example1} illustrates the evaluation steps of this program.

\begin{table*}
  \centering
  \caption{The execution steps of $S$. Each row contains the running process at each step and changes of communication qubits and heap immediately after evaluating the process.}
  \label{tbl:example1}
  \begin{tabular}{lccc}
    \toprule
    Running Process & $E$ (after, changes only) & $H$ (after, changes only) \\
    \midrule
    initial state
    & $\inquirmv{p_{{\mathrm{1}}}} \mapsto \{\overline{q}_{{\mathrm{1}}}\}, \inquirmv{p_{{\mathrm{2}}}} \mapsto \{\overline{q}_{{\mathrm{2}}}  \inquirsym{,}  \overline{q}_{{\mathrm{3}}}\}, \inquirmv{p_{{\mathrm{3}}}} \mapsto \{\overline{q}_{{\mathrm{4}}}\}$
    & $\emptyset$ \\
    \midrule
    $ \llbracket  \inquirmv{s}  \inquirsym{=} \, \inquirkw{open} \, \inquirsym{[}  \inquirmv{p_{{\mathrm{1}}}}  \inquirsym{,}  \inquirmv{p_{{\mathrm{2}}}}  \inquirsym{,}  \inquirmv{p_{{\mathrm{3}}}}  \inquirsym{]}  \rrbracket_{ \inquirmv{p_{{\mathrm{1}}}} }   \pipe \, .. \, \pipe   \llbracket  \inquirmv{s}  \inquirsym{=} \, \inquirkw{open} \, \inquirsym{[}  \inquirmv{p_{{\mathrm{1}}}}  \inquirsym{,}  \inquirmv{p_{{\mathrm{2}}}}  \inquirsym{,}  \inquirmv{p_{{\mathrm{3}}}}  \inquirsym{]}  \rrbracket_{ \inquirmv{p_{{\mathrm{3}}}} } $
    &
    & $ ( \inquirmv{s} ,  \inquirmv{p_{{\mathrm{1}}}} ) \mapsto   \epsilon  ,  ( \inquirmv{s} ,  \inquirmv{p_{{\mathrm{2}}}} ) \mapsto   \epsilon  ,  ( \inquirmv{s} ,  \inquirmv{p_{{\mathrm{3}}}} ) \mapsto   \epsilon  $ \\
    \midrule
    $ \llbracket  \inquirmv{x_{{\mathrm{2}}}}  \inquirsym{=} \, \inquirkw{genEnt} \, \inquirsym{[}  \inquirmv{p_{{\mathrm{2}}}}  \inquirsym{]}  \inquirsym{(}  \inquirmv{l_{{\mathrm{1}}}}  \inquirsym{)}  \rrbracket_{ \inquirmv{p_{{\mathrm{1}}}} }   \pipe   \llbracket  \inquirmv{z_{{\mathrm{1}}}}  \inquirsym{=} \, \inquirkw{genEnt} \, \inquirsym{[}  \inquirmv{p_{{\mathrm{1}}}}  \inquirsym{]}  \inquirsym{(}  \inquirmv{l_{{\mathrm{1}}}}  \inquirsym{)}  \rrbracket_{ \inquirmv{p_{{\mathrm{2}}}} } $
    & $\inquirmv{p_{{\mathrm{1}}}} \mapsto \emptyset, \inquirmv{p_{{\mathrm{2}}}} \mapsto \{\overline{q}_{{\mathrm{3}}}\}$
    & \\
    \midrule
    $ \llbracket  \inquirmv{z_{{\mathrm{2}}}}  \inquirsym{=} \, \inquirkw{genEnt} \, \inquirsym{[}  \inquirmv{p_{{\mathrm{3}}}}  \inquirsym{]}  \inquirsym{(}  \inquirmv{l_{{\mathrm{2}}}}  \inquirsym{)}  \rrbracket_{ \inquirmv{p_{{\mathrm{2}}}} }   \pipe   \llbracket  \inquirmv{y_{{\mathrm{1}}}}  \inquirsym{=} \, \inquirkw{genEnt} \, \inquirsym{[}  \inquirmv{p_{{\mathrm{2}}}}  \inquirsym{]}  \inquirsym{(}  \inquirmv{l_{{\mathrm{2}}}}  \inquirsym{)}  \rrbracket_{ \inquirmv{p_{{\mathrm{3}}}} } $
    & $\inquirmv{p_{{\mathrm{2}}}} \mapsto \emptyset, \inquirmv{p_{{\mathrm{3}}}} \mapsto \emptyset$
    & \\
    \midrule
    \begin{tabular}{l}
      $ \llbracket  \inquirsym{(}  \inquirmv{w_{{\mathrm{1}}}}  \inquirsym{,}  \inquirmv{w_{{\mathrm{2}}}}  \inquirsym{)}  \inquirsym{=} \, \inquirkw{entSwap} \, \inquirsym{(}  \overline{q}_{{\mathrm{2}}}  \inquirsym{,}  \overline{q}_{{\mathrm{3}}}  \inquirsym{)}  \rrbracket_{ \inquirmv{p_{{\mathrm{2}}}} } $ \\
      (the outcomes $\inquirmv{w_{{\mathrm{1}}}} = \inquirmv{v_{{\mathrm{1}}}}, \inquirmv{w_{{\mathrm{2}}}} = \inquirmv{v_{{\mathrm{2}}}}$)
    \end{tabular}
    & $\inquirmv{p_{{\mathrm{2}}}} \mapsto \{\overline{q}_{{\mathrm{2}}}  \inquirsym{,}  \overline{q}_{{\mathrm{3}}}\}$
    & \\
    \midrule
    $ \llbracket   \inquirmv{s} [  \inquirmv{p_{{\mathrm{1}}}}  ]!( \inquirmv{l_{{\mathrm{1}}}}  :  \inquirmv{v_{{\mathrm{1}}}} )   \rrbracket_{ \inquirmv{p_{{\mathrm{2}}}} } $ & & $ ( \inquirmv{s} ,  \inquirmv{p_{{\mathrm{1}}}} ) \mapsto  \inquirmv{l_{{\mathrm{1}}}}  \inquirsym{:}  \inquirmv{v_{{\mathrm{1}}}} $ \\
    \midrule
    $ \llbracket   \inquirmv{s} [  \inquirmv{p_{{\mathrm{3}}}}  ]!( \inquirmv{l_{{\mathrm{2}}}}  :  \inquirmv{v_{{\mathrm{2}}}} )   \rrbracket_{ \inquirmv{p_{{\mathrm{2}}}} } $ & & $ ( \inquirmv{s} ,  \inquirmv{p_{{\mathrm{3}}}} ) \mapsto  \inquirmv{l_{{\mathrm{2}}}}  \inquirsym{:}  \inquirmv{v_{{\mathrm{2}}}} $ \\
    \midrule
    $ \llbracket  \inquirmv{s}  \inquirsym{\mbox{?}}  \inquirsym{(}  \inquirmv{l_{{\mathrm{1}}}}  \inquirsym{:}  \inquirmv{w_{{\mathrm{1}}}}  \inquirsym{)}  \rrbracket_{ \inquirmv{p_{{\mathrm{1}}}} }   \pipe   \llbracket  \inquirmv{s}  \inquirsym{\mbox{?}}  \inquirsym{(}  \inquirmv{l_{{\mathrm{2}}}}  \inquirsym{:}  \inquirmv{w_{{\mathrm{2}}}}  \inquirsym{)}  \rrbracket_{ \inquirmv{p_{{\mathrm{3}}}} } $ & & $ ( \inquirmv{s} ,  \inquirmv{p_{{\mathrm{1}}}} ) \mapsto   \epsilon  ,  ( \inquirmv{s} ,  \inquirmv{p_{{\mathrm{3}}}} ) \mapsto   \epsilon  $ \\
    \midrule
    $ \llbracket    Z  ^{ \inquirmv{v_{{\mathrm{1}}}} }   \inquirsym{(}  \overline{q}_{{\mathrm{1}}}  \inquirsym{)}  \rrbracket_{ \inquirmv{p_{{\mathrm{1}}}} }   \pipe   \llbracket    X  ^{ \inquirmv{v_{{\mathrm{2}}}} }   \inquirsym{(}  \overline{q}_{{\mathrm{4}}}  \inquirsym{)}  \rrbracket_{ \inquirmv{p_{{\mathrm{3}}}} } $ & & \\
    \midrule
    $ \llbracket  \inquirkw{rcxc} \, \inquirsym{[}  \inquirmv{p_{{\mathrm{3}}}}  \inquirsym{]}  \inquirsym{(}  \inquirmv{s}  \inquirsym{,}  \inquirmv{l_{{\mathrm{3}}}}  \inquirsym{,}  q_{{\mathrm{1}}}  \inquirsym{,}  \overline{q}_{{\mathrm{1}}}  \inquirsym{)}  \rrbracket_{ \inquirmv{p_{{\mathrm{1}}}} }   \pipe   \llbracket  \inquirkw{rcxt} \, \inquirsym{[}  \inquirmv{p_{{\mathrm{1}}}}  \inquirsym{]}  \inquirsym{(}  \inquirmv{s}  \inquirsym{,}  \inquirmv{l_{{\mathrm{3}}}}  \inquirsym{,}  q_{{\mathrm{2}}}  \inquirsym{,}  \overline{q}_{{\mathrm{4}}}  \inquirsym{)}  \rrbracket_{ \inquirmv{p_{{\mathrm{3}}}} } $
    & $\inquirmv{p_{{\mathrm{1}}}} \mapsto \{\overline{q}_{{\mathrm{1}}}\}, \inquirmv{p_{{\mathrm{3}}}} \mapsto \{\overline{q}_{{\mathrm{4}}}\}$
    & \\
    \bottomrule
  \end{tabular}
\end{table*}

\begin{example}{(Barriers)}
  You may want to run a part of InQuIR processes synchronously even though InQuIR's processes work asynchronously.
  For example, imagine that all processors synchronize with each other for every instruction.
  InQuIR can encode synchronous communication with its asynchronous primitives.
  We show this by implementing $\inquirkw{barrier} \, \inquirsym{[}  p_{{\mathrm{1}}}  \inquirsym{,}  p_{{\mathrm{2}}}  \inquirsym{]}  \inquirsym{(}  \inquirmv{s}  \inquirsym{)}$, a barrier operation between processors $\inquirmv{p_{{\mathrm{1}}}}$ and $\inquirmv{p_{{\mathrm{2}}}}$ on a session $\inquirmv{s}$. The case of three or more processors can be implemented similarly.
  \begin{align*}
    & \llbracket  \inquirkw{barrier} \, \inquirsym{[}  \inquirmv{p_{{\mathrm{1}}}}  \inquirsym{,}  \inquirmv{p_{{\mathrm{2}}}}  \inquirsym{]}  \inquirsym{(}  \inquirmv{s}  \inquirsym{)}  \inquirsym{;}  \inquirmv{P_{{\mathrm{1}}}}  \rrbracket_{ \inquirmv{p_{{\mathrm{1}}}} }   \pipe   \llbracket  \inquirkw{barrier} \, \inquirsym{[}  \inquirmv{p_{{\mathrm{1}}}}  \inquirsym{,}  \inquirmv{p_{{\mathrm{2}}}}  \inquirsym{]}  \inquirsym{(}  \inquirmv{s}  \inquirsym{)}  \inquirsym{;}  \inquirmv{P_{{\mathrm{2}}}}  \rrbracket_{ \inquirmv{p_{{\mathrm{2}}}} }  \\
    &\quad\defeq  \llbracket   \inquirmv{s} [  p_{{\mathrm{2}}}  ]!( \inquirmv{l}  :  \inquirmv{v} )   \inquirsym{;}  \inquirmv{s}  \inquirsym{\mbox{?}}  \inquirsym{(}  \inquirmv{l}  \inquirsym{:}  \inquirmv{x_{{\mathrm{1}}}}  \inquirsym{)}  \inquirsym{;}  \inquirmv{P_{{\mathrm{1}}}}  \rrbracket_{ \inquirmv{p_{{\mathrm{1}}}} }  \\
    &\qquad\qquad\qquad\qquad \pipe  \llbracket  \inquirmv{s}  \inquirsym{\mbox{?}}  \inquirsym{(}  \inquirmv{l}  \inquirsym{:}  \inquirmv{x_{{\mathrm{2}}}}  \inquirsym{)}  \inquirsym{;}   \inquirmv{s} [  p_{{\mathrm{1}}}  ]!( \inquirmv{l}  :  \inquirmv{x_{{\mathrm{2}}}} )   \inquirsym{;}  \inquirmv{P_{{\mathrm{2}}}}  \rrbracket_{ \inquirmv{p_{{\mathrm{2}}}} } 
  \end{align*}
  , where $\inquirmv{l}$, $\inquirmv{v}$ and $\inquirmv{x_{\inquirmv{i}}}$ are dummies.
  We can express the blocking version of any non-blocking operations in the same way.
\end{example}

\begin{example}{(Qubit exhaustion)}\label{example:qubit-exhaust}
  The next example illustrates the movement of a data qubit by quantum teleportation.
  \begin{align*}
    \inquirmv{P_{{\mathrm{1}}}} &\defeq \inquirmv{x_{{\mathrm{1}}}}  \inquirsym{=} \, \inquirkw{genEnt} \, \inquirsym{[}  \inquirmv{p_{{\mathrm{2}}}}  \inquirsym{]}  \inquirsym{(}  \inquirmv{l}  \inquirsym{)}; \inquirkw{qsend} \, \inquirsym{[}  \inquirmv{p_{{\mathrm{2}}}}  \inquirsym{]}  \inquirsym{(}  \inquirmv{s}  \inquirsym{,}  \inquirmv{l}  \inquirsym{,}  \inquirmv{q_{{\mathrm{1}}}}  \inquirsym{,}  \inquirmv{x_{{\mathrm{1}}}}  \inquirsym{)} \\
    \inquirmv{P_{{\mathrm{2}}}} &\defeq \inquirmv{x_{{\mathrm{2}}}}  \inquirsym{=} \, \inquirkw{genEnt} \, \inquirsym{[}  \inquirmv{p_{{\mathrm{1}}}}  \inquirsym{]}  \inquirsym{(}  \inquirmv{l}  \inquirsym{)}; \\
    &\qquad \inquirmv{x_{{\mathrm{3}}}}  \inquirsym{=} \, \inquirkw{qrecv} \, \inquirsym{(}  \inquirmv{s}  \inquirsym{,}  \inquirmv{l}  \inquirsym{,}  \inquirmv{x_{{\mathrm{2}}}}  \inquirsym{)};  \mathit{CX}   \inquirsym{(}  \inquirmv{q_{{\mathrm{2}}}}  \inquirsym{,}  \inquirmv{x_{{\mathrm{3}}}}  \inquirsym{)} \\
    S &\defeq  \llbracket  \inquirmv{P_{{\mathrm{1}}}}  \rrbracket_{ \inquirmv{p_{{\mathrm{1}}}} }   \pipe   \llbracket  \inquirmv{P_{{\mathrm{2}}}}  \rrbracket_{ \inquirmv{p_{{\mathrm{2}}}} } 
  \end{align*}
\end{example}
The program $S$ applies a CX gate to remote qubits $\inquirmv{q_{{\mathrm{1}}}}$ and $\inquirmv{q_{{\mathrm{2}}}}$ by transfering $\inquirmv{q_{{\mathrm{1}}}}$ to the processor $\inquirmv{p_{{\mathrm{2}}}}$ and applying a local CX gate.
Unlike the remote $ \mathit{CX} $ operation, the operation $\mathbf{qsend}$ requires the receiver $\inquirmv{p_{{\mathrm{2}}}}$ to prepare a new qubit to receive the teleported quantum data.
If all data qubits are used in the processor $\inquirmv{p_{{\mathrm{2}}}}$ (that is, $Q(\inquirmv{p_{{\mathrm{2}}}} = \emptyset$)), the program will fail to receive the qubit.
In other words, the program will reach a stuck state.
We call this problem \textit{qubit exhaustion}.

\begin{example}{(Deadlock)}\label{example:deadlock}
  The final example shows a more practical issue called \textit{deadlock} caused by circular dependencies in a concurrent program.
  \begin{align*}
    \inquirmv{E} &\defeq \{\inquirmv{p_{{\mathrm{1}}}} \mapsto \{\overline{q}_{{\mathrm{1}}}\}, \inquirmv{p_{{\mathrm{2}}}} \mapsto \{\overline{q}_{{\mathrm{2}}}  \inquirsym{,}  \overline{q}_{{\mathrm{3}}}\}, \inquirmv{p_{{\mathrm{3}}}} \mapsto \{\overline{q}_{{\mathrm{4}}}\}\} \\
    \inquirmv{P_{{\mathrm{1}}}} &\defeq  \inquirmv{x_{{\mathrm{1}}}}  \inquirsym{=} \, \inquirkw{genEnt} \, \inquirsym{[}  \inquirmv{p_{{\mathrm{2}}}}  \inquirsym{]}  \inquirsym{(}  \inquirmv{l_{{\mathrm{1}}}}  \inquirsym{)} ; \cdots  \\
    \inquirmv{P_{{\mathrm{2}}}} &\defeq \inquirmv{y_{{\mathrm{1}}}}  \inquirsym{=} \, \inquirkw{genEnt} \, \inquirsym{[}  \inquirmv{p_{{\mathrm{1}}}}  \inquirsym{]}  \inquirsym{(}  \inquirmv{l_{{\mathrm{1}}}}  \inquirsym{)}  \inquirsym{;}  \inquirmv{y_{{\mathrm{2}}}}  \inquirsym{=} \, \inquirkw{genEnt} \, \inquirsym{[}  \inquirmv{p_{{\mathrm{3}}}}  \inquirsym{]}  \inquirsym{(}  \inquirmv{l_{{\mathrm{2}}}}  \inquirsym{)}; \\
    &\ \, \quad  \inquirsym{(}  \inquirmv{w_{{\mathrm{1}}}}  \inquirsym{,}  \inquirmv{w_{{\mathrm{2}}}}  \inquirsym{)}  \inquirsym{=} \, \inquirkw{entSwap} \, \inquirsym{(}  \inquirmv{y_{{\mathrm{1}}}}  \inquirsym{,}  \inquirmv{y_{{\mathrm{2}}}}  \inquirsym{)} ; \cdots  \\
    \inquirmv{P_{{\mathrm{3}}}} &\defeq  \inquirmv{z_{{\mathrm{1}}}}  \inquirsym{=} \, \inquirkw{genEnt} \, \inquirsym{[}  \inquirmv{p_{{\mathrm{2}}}}  \inquirsym{]}  \inquirsym{(}  \inquirmv{l_{{\mathrm{2}}}}  \inquirsym{)} ; \cdots  \\
    S_{{\mathrm{1}}} &\defeq   \llbracket  \inquirmv{P_{{\mathrm{1}}}}  \rrbracket_{ \inquirmv{p_{{\mathrm{1}}}} }   \pipe   \llbracket  \inquirmv{P_{{\mathrm{2}}}}  \rrbracket_{ \inquirmv{p_{{\mathrm{2}}}} }   \,|\,   \llbracket  \inquirmv{P_{{\mathrm{3}}}}  \rrbracket_{ \inquirmv{p_{{\mathrm{3}}}} }   \\
    \inquirmv{P_{{\mathrm{4}}}} &\defeq  \inquirmv{x_{{\mathrm{2}}}}  \inquirsym{=} \, \inquirkw{genEnt} \, \inquirsym{[}  \inquirmv{p_{{\mathrm{2}}}}  \inquirsym{]}  \inquirsym{(}  \inquirmv{l_{{\mathrm{3}}}}  \inquirsym{)} ; \cdots  \\
    \inquirmv{P_{{\mathrm{5}}}} &\defeq \inquirmv{y_{{\mathrm{3}}}}  \inquirsym{=} \, \inquirkw{genEnt} \, \inquirsym{[}  \inquirmv{p_{{\mathrm{1}}}}  \inquirsym{]}  \inquirsym{(}  \inquirmv{l_{{\mathrm{3}}}}  \inquirsym{)}  \inquirsym{;}  \inquirmv{y_{{\mathrm{4}}}}  \inquirsym{=} \, \inquirkw{genEnt} \, \inquirsym{[}  \inquirmv{p_{{\mathrm{3}}}}  \inquirsym{]}  \inquirsym{(}  \inquirmv{l_{{\mathrm{4}}}}  \inquirsym{)}; \\
    &\ \, \quad  \inquirsym{(}  \inquirmv{w_{{\mathrm{3}}}}  \inquirsym{,}  \inquirmv{w_{{\mathrm{4}}}}  \inquirsym{)}  \inquirsym{=} \, \inquirkw{entSwap} \, \inquirsym{(}  \inquirmv{y_{{\mathrm{3}}}}  \inquirsym{,}  \inquirmv{y_{{\mathrm{4}}}}  \inquirsym{)} ; \cdots  \\
    \inquirmv{P_{{\mathrm{6}}}} &\defeq  \inquirmv{z_{{\mathrm{2}}}}  \inquirsym{=} \, \inquirkw{genEnt} \, \inquirsym{[}  \inquirmv{p_{{\mathrm{2}}}}  \inquirsym{]}  \inquirsym{(}  \inquirmv{l_{{\mathrm{4}}}}  \inquirsym{)} ; \cdots  \\
    S_{{\mathrm{2}}} &\defeq   \llbracket  \inquirmv{P_{{\mathrm{4}}}}  \rrbracket_{ \inquirmv{p_{{\mathrm{1}}}} }   \pipe   \llbracket  \inquirmv{P_{{\mathrm{5}}}}  \rrbracket_{ \inquirmv{p_{{\mathrm{2}}}} }   \,|\,   \llbracket  \inquirmv{P_{{\mathrm{6}}}}  \rrbracket_{ \inquirmv{p_{{\mathrm{3}}}} }   \\
    S &\defeq S_{{\mathrm{1}}}  \pipe  S_{{\mathrm{2}}}
  \end{align*}
\end{example}
The systems $S_{{\mathrm{1}}}$ and $S_{{\mathrm{2}}}$ try entanglement swapping between the processors $\inquirmv{p_{{\mathrm{1}}}}$ and $\inquirmv{p_{{\mathrm{3}}}}$ in the same way.
We can interpret $S$ as a program where two different distributed programs $S_{{\mathrm{1}}}$ and $S_{{\mathrm{2}}}$ are running on the same interconnected quantum computers.
The execution order of $\inquirmv{P_{\inquirmv{i}}}\ (i = 1, \dots, 6)$ is not specified, so $S$ may proceed its evaluation like the following execution steps:
\begin{enumerate}
\item $ \llbracket  \inquirmv{P_{{\mathrm{1}}}}  \rrbracket_{ \inquirmv{p_{{\mathrm{1}}}} } $ and $ \llbracket  \inquirmv{P_{{\mathrm{2}}}}  \rrbracket_{ \inquirmv{p_{{\mathrm{2}}}} } $ generate an EPR pair and remove $\overline{q}_{{\mathrm{1}}}$ from $\inquirmv{E}$.
\item $ \llbracket  \inquirmv{P_{{\mathrm{5}}}}  \rrbracket_{ \inquirmv{p_{{\mathrm{2}}}} } $ and $ \llbracket  \inquirmv{P_{{\mathrm{6}}}}  \rrbracket_{ \inquirmv{p_{{\mathrm{3}}}} } $ generate an EPR pair and remove $\overline{q}_{{\mathrm{4}}}$ from $\inquirmv{E}$.
\item Try to generate an EPR pair between $\inquirmv{P_{{\mathrm{2}}}}$ and $\inquirmv{P_{{\mathrm{3}}}}$ (or between $\inquirmv{P_{{\mathrm{4}}}}$ and $\inquirmv{P_{{\mathrm{5}}}}$)
  to create a chain for entanglement swapping.
\end{enumerate}
Before the step 3, the exactly one communication qubit $\overline{q}_{{\mathrm{1}}}$ between $\inquirmv{p_{{\mathrm{1}}}}$ and $\inquirmv{p_{{\mathrm{2}}}}$ is consumed by $S_{{\mathrm{1}}}$,
and thus the system $S_{{\mathrm{2}}}$ has to wait until $S_{{\mathrm{1}}}$ returns it to $\inquirmv{E}$.
Similarly, $S_{{\mathrm{1}}}$ has to wait until $S_{{\mathrm{2}}}$ returns $\overline{q}_{{\mathrm{4}}}$ to $\inquirmv{E}$.
Both of $S_{{\mathrm{1}}}$ and $S_{{\mathrm{2}}}$ wait until each other's process is complete, and they become stuck.

We will discuss in \cref{sec:verification} how to deal with issues like \cref{example:qubit-exhaust} and \cref{example:deadlock}.

\section{Resource Analysis of InQuIR Programs}\label{sec:compiler}

In this section, we show that InQuIR can be used as a target language for distributed quantum compilers to estimate the performance of their compiled programs. We show this by implementing software tools, including a compiler and resource analyzer of InQuIR, and taking benchmarks on several quantum circuits under a variety of conditions. 

\subsection{Compilation Strategy}
Since one of the heaviest operations in distributed quantum computing is entanglement generation,
quantum compilers have to arrange the positions of data qubits to make the number of quantum communication as small as possible.
Since the development of a novel optimization method is out of the scope of this paper, we compile quantum programs based on the straightforward strategy that
\begin{itemize}
\item assigns qubits to quantum processors sequentially, and
\item translates any remote $\mathit{CX}$ into $\mathbf{rcxc}$ and $\mathbf{rcxt}$ along with entanglement swapping.
\end{itemize}
Note that InQuIR does not depend on a specific compiler, while our tools provide the toy compiler.

\subsection{Evaluation Method}

In this study, we calculated the cost of quantum programs based on various metrics. Specifically, this paper shows the following indicators using the InQuIR simulator:
\begin{itemize}
\item E-count, E-depth,
\item C-count, C-depth,
\item estimated time, and
\item the number of remaining operations by processors at each time.
\end{itemize}

E-count is the number of entanglement generations. E-depth is the length of the critical path when considering only the dependencies of entanglement generation. These metrics are often used in quantum interconnects because the cost of entanglement generation is significantly high. Similarly, C-count and C-depth are indicators for estimating the cost of classical communications.

\begin{remark}
Our E-depth calculation method is slightly different from the common method. In the quantum circuit representation often used in the analysis, the qubits are represented as wires, so it is clear which specific qubit is used by each instruction. InQuIR, on the other hand, does not specify which qubit is used in the initialization of qubits or the generation of entanglement, and the choice of qubits can change the result of the E-depth calculation. When multiple options are available, our simulator performs calculations based on the strategy of using the oldest qubit.
\end{remark}

The execution cost is an estimate of the execution time to complete processing on all processors. 
Here we use the following values for each operation based on the recent experimental data~\cite{sung2021RealizationHighFidelityCZ, sunada2022FastReadoutReset} and theoretical proposals for the heralded entanglement generation~\cite{barz2010heralded} scheme using transmon qubits and microwave-to-optical (M2O) converters~\cite{ang2022ArchitecturesMultinodeSuperconducting}.
\begin{itemize}
\item single-qubit gates: $\SI{30}{ns}$
\item local CX gate: $\SI{60}{ns}$
\item measurements: $\SI{240}{ns}$
\item classical communication: $\SI{30}{ns}$
\item entanglement generations: $\SI{1000}{ns}$
\end{itemize}
According to the need, these values can be freely customized per processor by supplying a JSON file.
We can analyze heterogeneous architectures by setting different costs for different processors.

The number of remaining operations by processors at each time allows us to analyze how efficiently each processor consumes their processes. Our tools visualize its time variation from timestamps that the simulator records when each operation is executed. This indicator allows analysis of the number of tasks assigned to processors and how busy they are at each time.

We calculated the computational costs described so far on the following architectures with various network topologies between processors:
\begin{itemize}
\item linearly connected processors,
\item (3D) cube, and
\item 2D torus.
\end{itemize}

We conducted experiments using small monolithic quantum circuits, which are widely used to validate quantum compilers~\cite{zulehner2018EfficientMethodologyMapping,li2022QASMBenchLowlevelQASM}\footnote{The dataset in the literature~\cite{zulehner2018EfficientMethodologyMapping} originated in RevLib~\cite{WGT+:2008}}.

\subsection{Evaluation Results}

\begin{table*}
  \centering
  \caption{The evaluation results. We used part of the dataset given in \cite{zulehner2018EfficientMethodologyMapping,li2022QASMBenchLowlevelQASM} to conduct the experiment. We compiled the dataset on three different topologies of processors: linear, 3D-cube, and torus. We write $(Q, E) \times M$ for the architecture with $M$ processors that have $Q$ data qubits and $E$ communication qubits. The symbol $D_E$ and $D_C$ denotes E-depth and C-depth, respectively. $N$ is the number of qubits declared in the program.}
  \label{tbl:eval-results}
  \begin{tabular}{lrrrrrrrrrrrr}
    \toprule
    \multirow{2}{*}{Circuit name} & \multirow{2}{*}{$N$} & \multirow{2}{*}{E-count} & \multirow{2}{*}{C-count} & \multicolumn{3}{c}{Linear: $(2, 2)\times 8$}  & \multicolumn{3}{c}{Linear: $(2, 4) \times 8$} & \multicolumn{3}{c}{Linear: $(2, 6) \times 8$} \\ \cmidrule(lr){5-7}\cmidrule(lr){8-10}\cmidrule(lr){11-13}
    & & & & $D_E$ & $D_C$ & cost$\mathrm{[ns]}$ & $D_E$ & $D_C$ & cost & $D_E$ & $D_C$ & cost $\mathrm{[ns]} $\\
    \midrule
    adr4\_197 & 16 & 5308 & 10616 & 1020 & 3150 & 1562850 & 510 & 2248 & 751630 & 510 & 2248 & 751630 \\
    ising\_model\_16 & 16 & 140 & 280 & 10 & 20 & 13510 & 5 & 20 & 7280 & 5 & 20 & 7280 \\
    rd53\_138 & 16 & 122 & 244 & 33 & 74 & 47730 & 17 & 66 & 23610 & 17 & 66 & 23610 \\
    sqn\_258 & 16 & 15054 & 30108 & 2843 & 9606 & 4393910 & 1365 & 6024 & 2136340 & 1365 & 6024 & 2136340 \\
    root\_255 & 16 & 31286 & 62572 & 5112 & 19268 & 8086560 & 2596 & 12878 & 3942970 & 2596 & 12878 & 3942970 \\
    4gt12-v1\_89 & 16 & 224 & 448 & 68 & 136 & 97370 & 34 & 118 & 48780 & 34 & 118 & 48780 \\
    9symml\_195 & 16 & 66732 & 133464 & 10512 & 40366 & 16504980 & 5342 & 25616 & 8025860 & 5342 & 25616 & 8025860 \\
    life\_238 & 16 & 42796 & 85592 & 6755 & 26076 & 10628990 & 3432 & 16564 & 5153840 & 3432 & 16564 & 5153840 \\
    \bottomrule
  \end{tabular}\vspace{5pt}
  
  \begin{tabular}{lrrrrrrrrrrr}
    \toprule
    \multirow{2}{*}{Circuit name} & \multirow{2}{*}{$N$} & \multicolumn{5}{c}{Cube: $(2, 3) \times 8$}  & \multicolumn{5}{c}{Torus: $(2, 4) \times 9$} \\ \cmidrule(lr){3-7}\cmidrule(lr){8-12}
    & & E-count & C-count & $D_E$ & $D_C$ & cost $\mathrm{[ns]}$ & E-count & C-count & $D_E$ & $D_C$ & cost $\mathrm{[ns]}$ \\
    \midrule
    adr4\_197 & 16 & 4300 & 8600 & 790 & 2214 & 1174120 & 3580 & 7160 & 574 & 1892 & 870770 \\
    ising\_model\_16 & 16 & 140 & 280 & 10 & 20 & 13510 & 180 & 360 & 10 & 22 & 14710  \\
    rd53\_138 & 16 & 122 & 244 & 31 & 72 & 44560 & 128 & 256 & 22 & 58 & 31870 \\
    sqn\_258 & 16 & 12238 & 24476 & 2104 & 6600 & 3269030 & 9762 & 19524 & 1933 & 5334 & 2922090 \\
    root\_255 & 16 & 22358 & 44716 & 3737 & 10858 & 5745630 & 18378 & 36756 & 2970 & 9324 & 4462430 \\
    4gt12-v1\_89 & 16 & 224 & 448 & 48 & 118 & 70680 & 152 & 304 & 48 & 104 & 66720 \\
    9symml\_195 & 16 & 50524 & 101048 & 7884 & 25472 & 11934160 & 39780 & 79560 & 8232 & 21696 & 12420720 \\
    life\_238 & 16 & 32484 & 64968 & 5039 & 16530 & 7675450 & 25408 & 50816 & 5073 & 13740 & 7675000 \\
    \bottomrule
  \end{tabular}
\end{table*}

\ifarxiv\else
{\xdef\xfigwd{\the\wd\figbox}
\fi
\begin{figure*}[tb]
  \begin{minipage}{0.5\hsize}
    \centering
    \includegraphics[width=9cm]{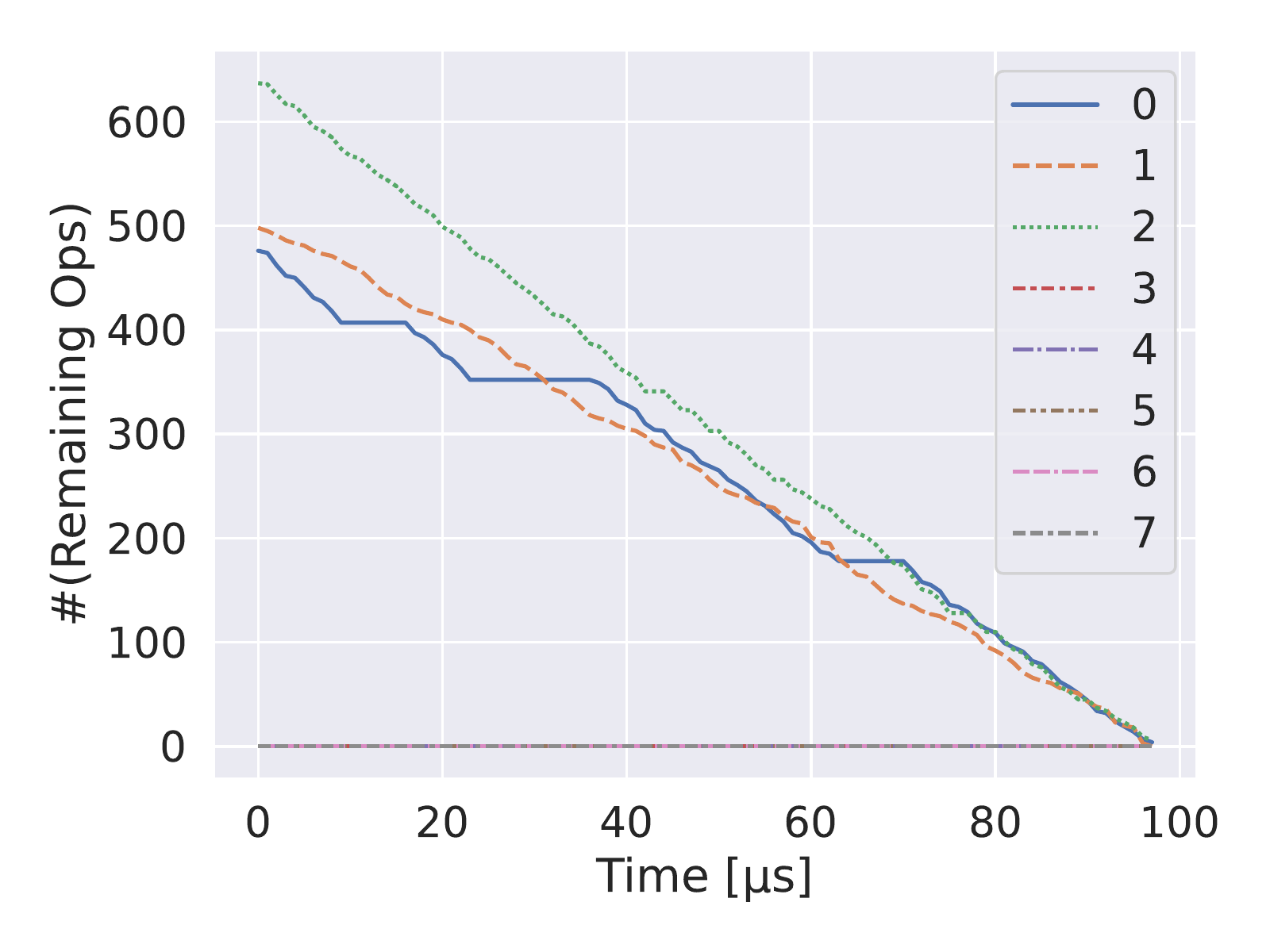}
  \end{minipage}
  \begin{minipage}{0.5\hsize}
    \centering
    \includegraphics[width=9cm]{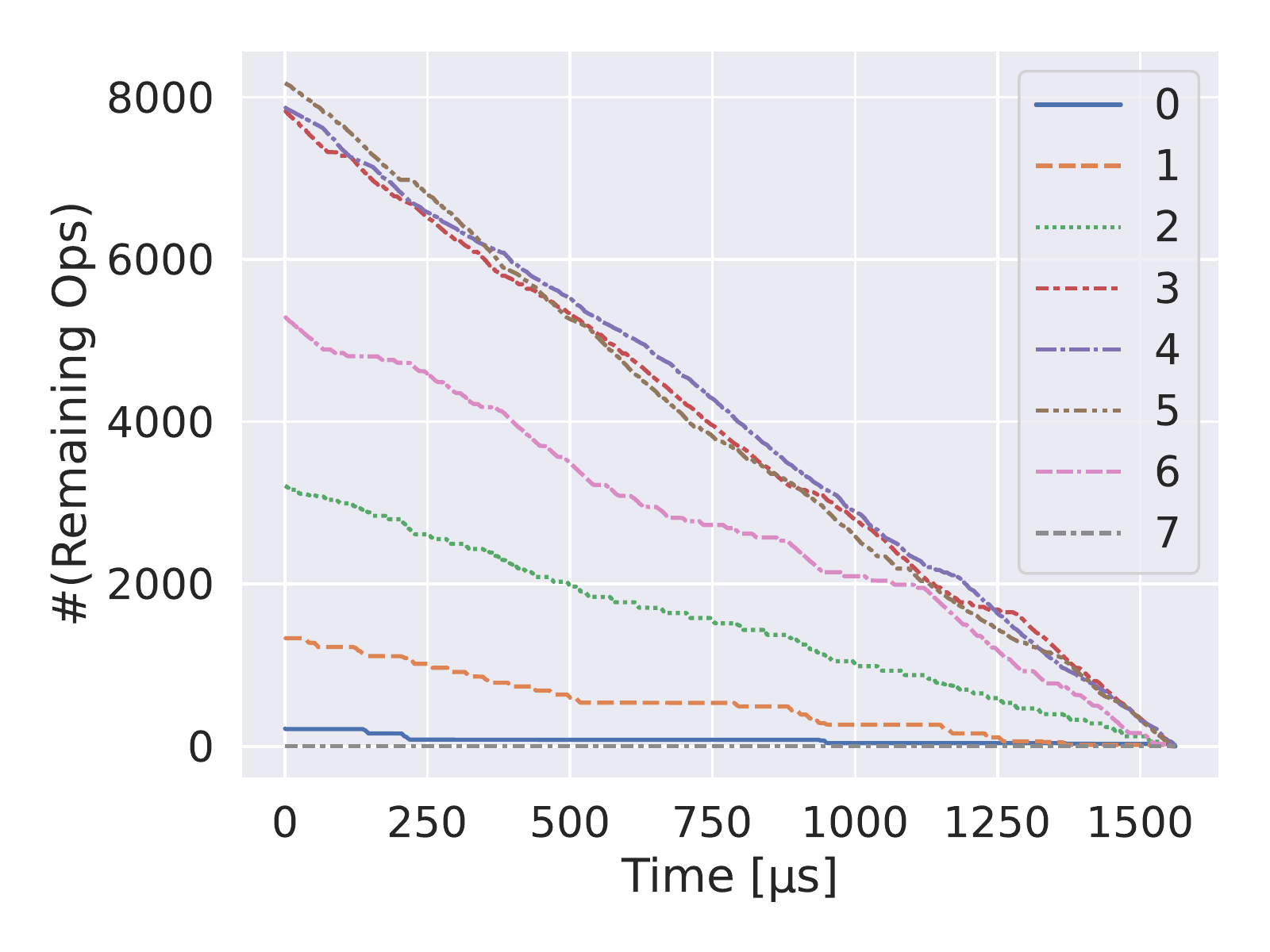}
  \end{minipage}
  \caption{The number of operations remaining at each time on each processor numbered from 0 to 7. The left and right graphs represent 4gt12-v1\_89 and adr4\_197, respectively. They were run on the $(2, 2) \times 8$ linearly connected architecture.}
  \label{fig:remaining-ops}
\end{figure*}
\ifarxiv\else
}
\fi

The detailed evaluation results are shown in \cref{tbl:eval-results} and \cref{fig:remaining-ops}. From here, we will explain what feedback to improve program performance can be obtained from the evaluation results.

In the upper table of \cref{tbl:eval-results}, we can see how computational costs change when the number of communication qubits is changed with linearly connected processors.
Here we denote $(Q, E) \times M$ for the architecture with $M$ processors with $Q$ data qubits and $E$ communication qubits.
From $(2, 2)\times 8$ to $(2, 4) \times 8$, each cost is almost halved. However, from $(2, 4) \times 8$ to $(2, 6) \times 8$, these value does not change. These results indicate how many communication qubits are required to optimize the computational cost.

In the lower table of \cref{tbl:eval-results}, we can see how the connectivity between processors influences the computational costs. It shows that the denser connectivity becomes, the smaller the computational cost is because the entanglement generation cost can be distributed. By comparing these values, we can choose the appropriate architecture for the algorithm we wish to run.

As shown in \cref{fig:remaining-ops}, we take 4gt12-v1\_89 and adr4\_197 circuits as examples to show how to obtain a hint to improve the compilation algorithm from the time variation of the number of remaining operations for each node.
In the case of 4gt12-v1\_89, it indicates that there are periods in which processor 0 is completely stopped during the calculation process. For example, we can see that it is stopped for $20$ to $\SI[parse-numbers=false]{40}{\micro\second}$ after starting the program execution.
In the case of adr4\_197, it can be seen that the number of operations assigned to each processor is very different.
These results suggest that there is room to reduce the total execution cost by improving the allocation of qubits to processors.
Otherwise, there will inherently be some qubits manipulated a lot and others not, which is another good hint for qubit allocation. For example, more frequently used qubits can be preferentially assigned to processors with higher performance.

Note that in both graphs, some processors do not execute any operations at all. This is because there are qubits that are declared in the program but not actually used.

As we have seen, we can use InQuIR to estimate program execution costs under various conditions. It is important to determine how much computational resources will be needed to run large quantum algorithms in the future. InQuIR can be a software foundation for the resource estimation of distributed quantum programs.

\section{Static Verification of InQuIR}\label{sec:verification}

As seen in \cref{sec:inquir}, an InQuIR program may get stuck during its execution.
The execution cost of a quantum program on actual devices is still high, and it is desirable to check whether a given program does not cause such errors without running it, that is, \emph{statically}.
InQuIR has formal semantics; thus, we can introduce static verification techniques, such as type systems and abstract interpretation.
We have not succeeded in constructing a complete verification system yet, so we will briefly provide how to guarantee that a given program is safe by a type system.

A type system's goal is to verify a given program's safety by proving the \textit{type safety}.
The type safety is usually written as follows: a well-typed program must not reach an \textit{undesirable} state.
An undesirable state depends on the case. For example, we can define an undesirable state as a stuck state.
We need to formalize a type system depending on the property we want to show.

In the case of InQuIR, a program can get stuck due to qubit exhaustion and deadlocks, as explained in \cref{example:qubit-exhaust} and \cref{example:deadlock}.
In order to avoid qubit exhaustion, we need quantum resource analysis that calculates how many qubits are used simultaneously.
We can analyze quantum resources with \textit{linear types}~\cite{turnerOnceType1995}, which is commonly used in many quantum programming languages to ensure that a qubit is consumed exactly once.
As for deadlock detection, we need dependency analysis between instructions in addition to the qubit resource analysis. For two operations $O_1$ and $O_2$, dependency analysis checks whether $O_1$ must be executed after $O_2$ or $O_1$ and $O_2$ may be performed simultaneously. Because InQuIR processes run asynchronously, this task is more complicated than in the case of a single process or synchronous processes. Fortunately, \textit{(multiparty) session types}~\cite{hondaMultipartyAsynchronousSession2008,hondakoheiMultipartyAsynchronousSession2016} have been studied widely to check the deadlock freedom of classical message passing processes. We believe that a similar type system will work well on InQuIR with a linear type system that tracks the consumptions of qubits and EPR pairs.

\section{Discussion and Future Work}\label{sec:discussion}

When running a distributed quantum program on a real quantum computer, a lot of information is unknown until runtime. For example, entanglement purification~\cite{bennett1996PurificationNoisyEntanglement, bennett1996mixed, bennett1996concentrating} is a stochastic procedure to generate a high-fidelity entanglement. Moreover, if a quantum program works for a long time, a computation node may be broken, or the error rate can become extremely high for some reason. In such a case, dynamic rescheduling is necessary to complete the program. Currently, the InQuIR assumes that a given program is scheduled statically, so there is room for extending the InQuIR with dynamic operations. Note that dynamic scheduling makes the behavior of distributed quantum programs more complicated, and a runtime error such as a deadlock may be more likely to occur.

The semantics of InQuIR adopts the policy that issues instructions in order from the front, basically the same as in classical computers. In many studies, however, instructions are issued right after their data dependencies have been resolved, and InQuIR also uses this policy to evaluate the computation cost of InQuIR programs.
The latter policy has the advantage that each instruction can be issued as soon as possible. At the same time, the runtime cost at each execution cycle is likely higher since it needs to maintain a dependency graph. It is difficult to say at this time which of the two approaches is better because the amount of time that can be spent on each execution cycle depends on a variety of factors, including the architecture, instruction set, and computational paradigm of the future quantum computer.

The quantum devices that exist at this time are Noisy Intermediate Scale Quantum (NISQ) devices where the effects of noise cannot be ignored. When trying to use InQuIR in such a system, it is necessary to compile it into a high-quality executable quantum circuit using the various constraints that the device has. In particular, to reduce noise during gate execution, the circuit must be compiled so that the number of gates is smaller, taking into account the qubit connectivity inside the processor\cite{holmes2020impact}. In addition, the accuracy of the gates on each qubit is non-uniform, so sensitive qubit assignment is necessary\cite{murali2019noise, nishio2020extracting}.

In order to realize large-scale quantum computation in the future, it is believed that fault-tolerant quantum computation (FTQC)~\cite{gottesman1997stabilizer} using quantum error-correcting codes will be necessary. The basic structure and analysis methods of InQuIR are general enough to be used in the design of FTQC languages for distributed quantum systems. However, the code structure, instruction set, microarchitecture, and compilation strategy need to be properly formulated\cite{li2016hierarchical}.

InQuIR supports the quantum analog of the No-remote memory access model (NORA) architectures and does not assume shared memory. However, suppose someone defines instruction sets that can handle quantum analogs of the Non-Uniform Access Memory model (NUMA) architectures~\cite{lameter2013numa}, including access to shared memory. In that case, InQuIR can be used as a compile target from such a program. 

\section{Conclusion}\label{sec:conclusion}
We proposed InQuIR, an intermediate representation to describe distributed quantum computation on multiple quantum processors and enhance the reusability of distributed quantum programs. We designed InQuIR as a formally defined programming language to discuss distributed quantum programs' behaviors precisely. We gave several examples written in InQuIR to show the realization of remote operations and undesirable runtime states such as deadlocks. We discussed a type system to detect such runtime errors before execution. Furthermore, we implemented software tools to evaluate the performance of programs written in InQuIR.

\section*{Acknowledgements}
SN and RW acknowledge Atsushi Igarashi and Kae Nemoto for useful discussions throughout this project.
This work was supported in part by JSPS KAKENHI Grant Number JP22J20882
and by JST, the establishment of university fellowships towards
the creation of science technology innovation, Grant Number JPMJFS2123.

\bibliographystyle{ieeetr}
\bibliography{main}

\begin{thebibliography}{10}

\bibitem{adedoyin2018quantum}
A.~Adedoyin, J.~Ambrosiano, P.~Anisimov, A.~B{\"a}rtschi, W.~Casper,
  G.~Chennupati, C.~Coffrin, H.~Djidjev, D.~Gunter, S.~Karra, {\em et~al.},
  ``Quantum algorithm implementations for beginners,'' {\em arXiv preprint
  arXiv:1804.03719}, 2018.

\bibitem{chiaverini2004realization}
J.~Chiaverini, D.~Leibfried, T.~Schaetz, M.~D. Barrett, R.~Blakestad,
  J.~Britton, W.~M. Itano, J.~D. Jost, E.~Knill, C.~Langer, {\em et~al.},
  ``Realization of quantum error correction,'' {\em Nature}, vol.~432,
  no.~7017, pp.~602--605, 2004.

\bibitem{krinner2019engineering}
S.~Krinner, S.~Storz, P.~Kurpiers, P.~Magnard, J.~Heinsoo, R.~Keller,
  J.~Luetolf, C.~Eichler, and A.~Wallraff, ``Engineering cryogenic setups for
  100-qubit scale superconducting circuit systems,'' {\em EPJ Quantum
  Technology}, vol.~6, no.~1, p.~2, 2019.

\bibitem{gold2021entanglement}
A.~Gold, J.~Paquette, A.~Stockklauser, M.~J. Reagor, M.~S. Alam, A.~Bestwick,
  N.~Didier, A.~Nersisyan, F.~Oruc, A.~Razavi, {\em et~al.}, ``Entanglement
  across separate silicon dies in a modular superconducting qubit device,''
  {\em npj Quantum Information}, vol.~7, no.~1, pp.~1--10, 2021.

\bibitem{tamate2021toward}
S.~Tamate, Y.~Tabuchi, and Y.~Nakamura, ``Toward realization of scalable
  packaging and wiring for large-scale superconducting quantum computers,''
  {\em IEICE Transactions on Electronics}, 2021.

\bibitem{Awschalom2021}
D.~Awschalom, K.~K. Berggren, H.~Bernien, S.~Bhave, L.~D. Carr, P.~Davids,
  S.~E. Economou, D.~Englund, A.~Faraon, M.~Fejer, S.~Guha, M.~V. Gustafsson,
  E.~Hu, L.~Jiang, J.~Kim, B.~Korzh, P.~Kumar, P.~G. Kwiat, M.~Lon^^c4^^8dar,
  M.~D. Lukin, D.~A. Miller, C.~Monroe, S.~W. Nam, P.~Narang, J.~S. Orcutt,
  M.~G. Raymer, A.~H. Safavi-Naeini, M.~Spiropulu, K.~Srinivasan, S.~Sun,
  J.~Vu^^c4^^8dkovi^^c4^^87, E.~Waks, R.~Walsworth, A.~M. Weiner, and Z.~Zhang,
  ``Development of quantum interconnects (quics) for next-generation
  information technologies,'' {\em PRX Quantum}, vol.~2, pp.~1--21, 2021.

\bibitem{ben2005fast}
M.~Ben-Or and A.~Hassidim, ``Fast quantum byzantine agreement,'' in {\em
  Proceedings of the thirty-seventh annual ACM symposium on Theory of
  computing}, pp.~481--485, 2005.

\bibitem{barz2012demonstration}
S.~Barz, E.~Kashefi, A.~Broadbent, J.~F. Fitzsimons, A.~Zeilinger, and
  P.~Walther, ``Demonstration of blind quantum computing,'' {\em science},
  vol.~335, no.~6066, pp.~303--308, 2012.

\bibitem{hanerDistributedQuantumComputing2021}
T.~H{\"a}ner, D.~S. Steiger, T.~Hoefler, and M.~Troyer, ``Distributed quantum
  computing with {{QMPI}},'' {\em Proceedings of the International Conference
  for High Performance Computing, Networking, Storage and Analysis}, pp.~1--13,
  Nov. 2021.

\bibitem{cuomoOptimizedCompilerDistributed2021}
D.~Cuomo, M.~Caleffi, K.~Krsulich, F.~Tramonto, G.~Agliardi, E.~Prati, and
  A.~S. Cacciapuoti, ``Optimized compiler for distributed quantum computing,''
  {\em arXiv:2112.14139 [quant-ph]}, Dec. 2021.

\bibitem{ferrariCompilerDesignDistributed2021}
D.~Ferrari, A.~S. Cacciapuoti, M.~Amoretti, and M.~Caleffi, ``Compiler design
  for distributed quantum computing,'' {\em IEEE Transactions on Quantum
  Engineering}, vol.~2, pp.~1--20, 2021.

\bibitem{meter2008arithmetic}
R.~V. Meter, W.~Munro, K.~Nemoto, and K.~M. Itoh, ``Arithmetic on a
  distributed-memory quantum multicomputer,'' {\em ACM Journal on Emerging
  Technologies in Computing Systems (JETC)}, vol.~3, no.~4, pp.~1--23, 2008.

\bibitem{mccaskeyMLIRDialectQuantum2021}
A.~McCaskey and T.~Nguyen, ``A {{MLIR Dialect}} for {{Quantum Assembly
  Languages}},'' {\em arXiv:2101.11365 [quant-ph]}, Jan. 2021.

\bibitem{frontier2021}
D.~R.~N. Laboratory, ``Frontier - hpe cray ex235a, amd optimized 3rd generation
  epyc 64c 2ghz, amd instinct mi250x, slingshot-11, hpe,'' tech. rep., USA.
\newblock \url{https://www.olcf.ornl.gov/frontier/}.

\bibitem{dongarra2020report}
J.~Dongarra, ``Report on the fujitsu fugaku system,'' {\em University of
  Tennessee-Knoxville Innovative Computing Laboratory, Tech. Rep. ICLUT-20-06},
  2020.

\bibitem{lumi2022}
EuroHPC/CSC, ``Lumi supercomputer,'' tech. rep., Finland.
\newblock
  \url{https://www.lumi-supercomputer.eu/lumis-fullsystem-architecture-revealed/}.

\bibitem{greenberger1989going}
D.~Greenberger, M.~Horne, and A.~Zeilinger, ``Going beyond bell’s theorem
  bell’s theorem, quantum theory and conceptions of the universe ed m
  kafatos,'' {\em Dordrecht: Kluwer}, vol.~69, pp.~69--72, 1989.

\bibitem{raussendorf2001one}
R.~Raussendorf and H.~J. Briegel, ``A one-way quantum computer,'' {\em Physical
  Review Letters}, vol.~86, no.~22, p.~5188, 2001.

\bibitem{raussendorf2003measurement}
R.~Raussendorf, D.~E. Browne, and H.~J. Briegel, ``Measurement-based quantum
  computation on cluster states,'' {\em Physical review A}, vol.~68, no.~2,
  p.~022312, 2003.

\bibitem{zomorodi2018optimizing}
M.~Zomorodi-Moghadam, M.~Houshmand, and M.~Houshmand, ``Optimizing
  teleportation cost in distributed quantum circuits,'' {\em International
  Journal of Theoretical Physics}, vol.~57, no.~3, pp.~848--861, 2018.

\bibitem{daei2021improving}
O.~Daei, K.~Navi, and M.~Zomorodi, ``Improving the teleportation cost in
  distributed quantum circuits based on commuting of gates,'' {\em
  International Journal of Theoretical Physics}, vol.~60, no.~9,
  pp.~3494--3513, 2021.

\bibitem{nikahd2021automated}
E.~Nikahd, N.~Mohammadzadeh, M.~Sedighi, and M.~S. Zamani, ``Automated
  window-based partitioning of quantum circuits,'' {\em Physica Scripta},
  vol.~96, no.~3, p.~035102, 2021.

\bibitem{dadkhah2021new}
D.~Dadkhah, M.~Zomorodi, and S.~E. Hosseini, ``A new approach for optimization
  of distributed quantum circuits,'' {\em International Journal of Theoretical
  Physics}, vol.~60, no.~9, pp.~3271--3285, 2021.

\bibitem{g2021efficient}
R.~G~Sundaram, H.~Gupta, and C.~Ramakrishnan, ``Efficient distribution of
  quantum circuits,'' in {\em 35th International Symposium on Distributed
  Computing (DISC 2021)}, Schloss Dagstuhl-Leibniz-Zentrum f{\"u}r Informatik,
  2021.

\bibitem{hammerer2010quantum}
K.~Hammerer, A.~S. S{\o}rensen, and E.~S. Polzik, ``Quantum interface between
  light and atomic ensembles,'' {\em Reviews of Modern Physics}, vol.~82,
  no.~2, p.~1041, 2010.

\bibitem{andrews2014bidirectional}
R.~W. Andrews, R.~W. Peterson, T.~P. Purdy, K.~Cicak, R.~W. Simmonds, C.~A.
  Regal, and K.~W. Lehnert, ``Bidirectional and efficient conversion between
  microwave and optical light,'' {\em Nature physics}, vol.~10, no.~4,
  pp.~321--326, 2014.

\bibitem{mirhosseini2020superconducting}
M.~Mirhosseini, A.~Sipahigil, M.~Kalaee, and O.~Painter, ``Superconducting
  qubit to optical photon transduction,'' {\em Nature}, vol.~588, no.~7839,
  pp.~599--603, 2020.

\bibitem{brubaker2022optomechanical}
B.~M. Brubaker, J.~M. Kindem, M.~D. Urmey, S.~Mittal, R.~D. Delaney, P.~S.
  Burns, M.~R. Vissers, K.~W. Lehnert, and C.~A. Regal, ``Optomechanical
  ground-state cooling in a continuous and efficient electro-optic
  transducer,'' {\em Physical Review X}, vol.~12, no.~2, p.~021062, 2022.

\bibitem{fan2018superconducting}
L.~Fan, C.-L. Zou, R.~Cheng, X.~Guo, X.~Han, Z.~Gong, S.~Wang, and H.~X. Tang,
  ``Superconducting cavity electro-optics: a platform for coherent photon
  conversion between superconducting and photonic circuits,'' {\em Science
  advances}, vol.~4, no.~8, p.~eaar4994, 2018.

\bibitem{sahu2022quantum}
R.~Sahu, W.~Hease, A.~Rueda, G.~Arnold, L.~Qiu, and J.~M. Fink,
  ``Quantum-enabled operation of a microwave-optical interface,'' {\em Nature
  communications}, vol.~13, no.~1, pp.~1--7, 2022.

\bibitem{fernandez2019cavity}
X.~Fernandez-Gonzalvo, S.~P. Horvath, Y.-H. Chen, and J.~J. Longdell,
  ``Cavity-enhanced raman heterodyne spectroscopy in er 3+: Y 2 sio 5 for
  microwave to optical signal conversion,'' {\em Physical Review A}, vol.~100,
  no.~3, p.~033807, 2019.

\bibitem{vogt2019efficient}
T.~Vogt, C.~Gross, J.~Han, S.~B. Pal, M.~Lam, M.~Kiffner, and W.~Li,
  ``Efficient microwave-to-optical conversion using rydberg atoms,'' {\em
  Physical Review A}, vol.~99, no.~2, p.~023832, 2019.

\bibitem{kimble1998strong}
H.~J. Kimble, ``Strong interactions of single atoms and photons in cavity
  qed,'' {\em Physica Scripta}, vol.~1998, no.~T76, p.~127, 1998.

\bibitem{girvin2009circuit}
S.~Girvin, M.~Devoret, and R.~Schoelkopf, ``Circuit qed and engineering
  charge-based superconducting qubits,'' {\em Physica Scripta}, vol.~2009,
  no.~T137, p.~014012, 2009.

\bibitem{gisin2007quantum}
N.~Gisin and R.~Thew, ``Quantum communication,'' {\em Nature photonics},
  vol.~1, no.~3, pp.~165--171, 2007.

\bibitem{kimble2008quantum}
H.~J. Kimble, ``The quantum internet,'' {\em Nature}, vol.~453, no.~7198,
  pp.~1023--1030, 2008.

\bibitem{gottesman2001encoding}
D.~Gottesman, A.~Kitaev, and J.~Preskill, ``Encoding a qubit in an
  oscillator,'' {\em Physical Review A}, vol.~64, no.~1, p.~012310, 2001.

\bibitem{michael2016new}
M.~H. Michael, M.~Silveri, R.~Brierley, V.~V. Albert, J.~Salmilehto, L.~Jiang,
  and S.~M. Girvin, ``New class of quantum error-correcting codes for a bosonic
  mode,'' {\em Physical Review X}, vol.~6, no.~3, p.~031006, 2016.

\bibitem{grassl1999quantum}
M.~Grassl, W.~Geiselmann, and T.~Beth, ``Quantum reed―solomon codes,'' in {\em
  International Symposium on Applied Algebra, Algebraic Algorithms, and
  Error-Correcting Codes}, pp.~231--244, Springer, 1999.

\bibitem{briegel1998quantum}
H.-J. Briegel, W.~D{\"u}r, J.~I. Cirac, and P.~Zoller, ``Quantum repeaters: the
  role of imperfect local operations in quantum communication,'' {\em Physical
  Review Letters}, vol.~81, no.~26, p.~5932, 1998.

\bibitem{van2014quantum}
R.~Van~Meter, {\em Quantum networking}.
\newblock John Wiley \& Sons, 2014.

\bibitem{wehner2018quantum}
S.~Wehner, D.~Elkouss, and R.~Hanson, ``Quantum internet: A vision for the road
  ahead,'' {\em Science}, vol.~362, no.~6412, p.~eaam9288, 2018.

\bibitem{kozlowski2020designing}
W.~Kozlowski, A.~Dahlberg, and S.~Wehner, ``Designing a quantum network
  protocol,'' in {\em Proceedings of the 16th International Conference on
  emerging Networking EXperiments and Technologies}, pp.~1--16, 2020.

\bibitem{bennett1993teleporting}
C.~H. Bennett, G.~Brassard, C.~Cr{\'e}peau, R.~Jozsa, A.~Peres, and W.~K.
  Wootters, ``Teleporting an unknown quantum state via dual classical and
  einstein-podolsky-rosen channels,'' {\em Physical review letters}, vol.~70,
  no.~13, p.~1895, 1993.

\bibitem{gottesman1999teleport}
D.~Gottesman and I.~L. Chuang, ``Quantum {{Teleportation}} is a {{Universal
  Computational Primitive}},'' {\em Nature}, vol.~402, pp.~390--393, Nov. 1999.

\bibitem{jiang2007distributed}
L.~Jiang, J.~M. Taylor, A.~S. S{\o}rensen, and M.~D. Lukin, ``Distributed
  quantum computation based on small quantum registers,'' {\em Physical Review
  A}, vol.~76, no.~6, p.~062323, 2007.

\bibitem{zukowski1993event}
M.~Zukowski, A.~Zeilinger, M.~A. Horne, and A.~K. Ekert, ``"
  event-ready-detectors" bell experiment via entanglement swapping.,'' {\em
  Physical Review Letters}, vol.~71, no.~26, 1993.

\bibitem{dahlbergNetQASMLowlevelInstruction2021}
A.~Dahlberg, B.~{van der Vecht}, C.~D. Donne, M.~Skrzypczyk, I.~te~Raa,
  W.~Kozlowski, and S.~Wehner, ``{{NetQASM}} -- {{A}} low-level instruction set
  architecture for hybrid quantum-classical programs in a quantum internet,''
  {\em arXiv:2111.09823 [quant-ph]}, Nov. 2021.

\bibitem{cross2017open}
A.~W. Cross, L.~S. Bishop, J.~A. Smolin, and J.~M. Gambetta, ``Open quantum
  assembly language,'' {\em arXiv preprint arXiv:1707.03429}, 2017.

\bibitem{gay2004CommunicatingQuantum}
S.~Gay and R.~Nagarajan, ``Communicating {{Quantum Processes}},'' Sept. 2004.

\bibitem{milner1992CalculusMobile}
R.~Milner, J.~Parrow, and D.~Walker, ``A calculus of mobile processes, {{I}},''
  {\em Information and Computation}, vol.~100, pp.~1--40, Sept. 1992.

\bibitem{gay2005ProbabilisticModel}
S.~Gay, R.~Nagarajan, and N.~Papanikolaou, ``Probabilistic {{Model--Checking}}
  of {{Quantum Protocols}},'' Oct. 2005.

\bibitem{nagarajan2005AutomatedAnalysis}
R.~Nagarajan, N.~Papanikolaou, G.~Bowen, and S.~Gay, ``An {{Automated
  Analysis}} of the {{Security}} of {{Quantum Key Distribution}},'' Feb. 2005.

\bibitem{hennessy2002ResourceAccessControl}
M.~Hennessy and J.~Riely, ``Resource {{Access Control}} in {{Systems}} of
  {{Mobile Agents}},'' {\em Information and Computation}, vol.~173,
  pp.~82--120, Feb. 2002.

\bibitem{dahlberg2018simulaqron}
A.~Dahlberg and S.~Wehner, ``Simulaqron―a simulator for developing quantum
  internet software,'' {\em Quantum Science and Technology}, vol.~4, no.~1,
  p.~015001, 2018.

\bibitem{bartlett2018distributed}
B.~Bartlett, ``A distributed simulation framework for quantum networks and
  channels,'' {\em arXiv preprint arXiv:1808.07047}, 2018.

\bibitem{satoh2021quisp}
R.~Satoh, M.~Hajdu{\v{s}}ek, N.~Benchasattabuse, S.~Nagayama, K.~Teramoto,
  T.~Matsuo, S.~A. Metwalli, T.~Satoh, S.~Suzuki, and R.~Van~Meter, ``Quisp: a
  quantum internet simulation package,'' {\em arXiv preprint arXiv:2112.07093},
  2021.

\bibitem{coopmans2021netsquid}
T.~Coopmans, R.~Knegjens, A.~Dahlberg, D.~Maier, L.~Nijsten,
  J.~de~Oliveira~Filho, M.~Papendrecht, J.~Rabbie, F.~Rozp{\k{e}}dek,
  M.~Skrzypczyk, {\em et~al.}, ``Netsquid, a network simulator for quantum
  information using discrete events,'' {\em Communications Physics}, vol.~4,
  no.~1, pp.~1--15, 2021.

\bibitem{wu2021sequence}
X.~Wu, A.~Kolar, J.~Chung, D.~Jin, T.~Zhong, R.~Kettimuthu, and M.~Suchara,
  ``Sequence: a customizable discrete-event simulator of quantum networks,''
  {\em Quantum Science and Technology}, vol.~6, no.~4, p.~045027, 2021.

\bibitem{diadamo2021qunetsim}
S.~DiAdamo, J.~N{\"o}tzel, B.~Zanger, and M.~M. Be{\c{s}}e, ``Qunetsim: A
  software framework for quantum networks,'' {\em IEEE Transactions on Quantum
  Engineering}, vol.~2, pp.~1--12, 2021.

\bibitem{danos2005DistributedMeasurementbasedQuantum}
V.~Danos, E.~D'Hondt, E.~Kashefi, and P.~Panangaden, ``Distributed
  measurement-based quantum computation,'' June 2005.

\bibitem{10.5555/898758}
M.~P.~I. Forum, ``Mpi: A message-passing interface standard,'' tech. rep., USA,
  1994.

\bibitem{bichsel2020SilqHighlevelQuantum}
B.~Bichsel, M.~Baader, T.~Gehr, and M.~Vechev, ``Silq: A high-level quantum
  language with safe uncomputation and intuitive semantics,'' in {\em
  Proceedings of the 41st {{ACM SIGPLAN Conference}} on {{Programming Language
  Design}} and {{Implementation}}}, {{PLDI}} 2020, ({New York, NY, USA}),
  pp.~286--300, {Association for Computing Machinery}, June 2020.

\bibitem{sung2021RealizationHighFidelityCZ}
Y.~Sung, L.~Ding, J.~Braum{\"u}ller, A.~Veps{\"a}l{\"a}inen, B.~Kannan,
  M.~Kjaergaard, A.~Greene, G.~O. Samach, C.~McNally, D.~Kim, A.~Melville,
  B.~M. Niedzielski, M.~E. Schwartz, J.~L. Yoder, T.~P. Orlando, S.~Gustavsson,
  and W.~D. Oliver, ``Realization of {{High-Fidelity CZ}} and \${{ZZ}}\$-{{Free
  iSWAP Gates}} with a {{Tunable Coupler}},'' {\em Physical Review X}, vol.~11,
  p.~021058, June 2021.

\bibitem{sunada2022FastReadoutReset}
Y.~Sunada, S.~Kono, J.~Ilves, S.~Tamate, T.~Sugiyama, Y.~Tabuchi, and
  Y.~Nakamura, ``Fast {{Readout}} and {{Reset}} of a {{Superconducting Qubit
  Coupled}} to a {{Resonator}} with an {{Intrinsic Purcell Filter}},'' {\em
  Physical Review Applied}, vol.~17, p.~044016, Apr. 2022.

\bibitem{barz2010heralded}
S.~Barz, G.~Cronenberg, A.~Zeilinger, and P.~Walther, ``Heralded generation of
  entangled photon pairs,'' {\em Nature photonics}, vol.~4, no.~8,
  pp.~553--556, 2010.

\bibitem{ang2022ArchitecturesMultinodeSuperconducting}
J.~Ang, G.~Carini, Y.~Chen, I.~Chuang, M.~A. DeMarco, S.~E. Economou,
  A.~Eickbusch, A.~Faraon, K.-M. Fu, S.~M. Girvin, M.~Hatridge, A.~Houck,
  P.~Hilaire, K.~Krsulich, A.~Li, C.~Liu, Y.~Liu, M.~Martonosi, D.~C. McKay,
  J.~Misewich, M.~Ritter, R.~J. Schoelkopf, S.~A. Stein, S.~Sussman, H.~X.
  Tang, W.~Tang, T.~Tomesh, N.~M. Tubman, C.~Wang, N.~Wiebe, Y.-X. Yao, D.~C.
  Yost, and Y.~Zhou, ``Architectures for {{Multinode Superconducting Quantum
  Computers}},'' Dec. 2022.

\bibitem{zulehner2018EfficientMethodologyMapping}
A.~Zulehner, A.~Paler, and R.~Wille, ``An {{Efficient Methodology}} for
  {{Mapping Quantum Circuits}} to the {{IBM QX Architectures}},'' June 2018.

\bibitem{li2022QASMBenchLowlevelQASM}
A.~Li, S.~Stein, S.~Krishnamoorthy, and J.~Ang, ``{{QASMBench}}: {{A Low-level
  QASM Benchmark Suite}} for {{NISQ Evaluation}} and {{Simulation}},'' May
  2022.

\bibitem{WGT+:2008}
R.~Wille, D.~Gro{\ss}e, L.~Teuber, G.~W. Dueck, and R.~Drechsler, ``{{RevLib}}:
  {{An}} online resource for reversible functions and reversible circuits,'' in
  {\em Int'l Symp. on Multi-Valued Logic}, pp.~220--225, 2008.

\bibitem{turnerOnceType1995}
D.~N. Turner, P.~Wadler, and C.~Mossin, ``Once upon a type,'' in {\em
  Proceedings of the Seventh International Conference on {{Functional}}
  Programming Languages and Computer Architecture}, {{FPCA}} '95, ({New York,
  NY, USA}), pp.~1--11, {Association for Computing Machinery}, Oct. 1995.

\bibitem{hondaMultipartyAsynchronousSession2008}
K.~Honda, N.~Yoshida, and M.~Carbone, ``Multiparty asynchronous session
  types,'' {\em ACM SIGPLAN Notices}, vol.~43, pp.~273--284, Jan. 2008.

\bibitem{hondakoheiMultipartyAsynchronousSession2016}
HondaKohei, YoshidaNobuko, and CarboneMarco, ``Multiparty {{Asynchronous
  Session Types}},'' {\em Journal of the ACM (JACM)}, Mar. 2016.

\bibitem{bennett1996PurificationNoisyEntanglement}
C.~H. Bennett, G.~Brassard, S.~Popescu, B.~Schumacher, J.~A. Smolin, and W.~K.
  Wootters, ``Purification of {{Noisy Entanglement}} and {{Faithful
  Teleportation}} via {{Noisy Channels}},'' {\em Physical Review Letters},
  vol.~76, pp.~722--725, Jan. 1996.

\bibitem{bennett1996mixed}
C.~H. Bennett, D.~P. DiVincenzo, J.~A. Smolin, and W.~K. Wootters,
  ``Mixed-state entanglement and quantum error correction,'' {\em Physical
  Review A}, vol.~54, no.~5, p.~3824, 1996.

\bibitem{bennett1996concentrating}
C.~H. Bennett, H.~J. Bernstein, S.~Popescu, and B.~Schumacher, ``Concentrating
  partial entanglement by local operations,'' {\em Physical Review A}, vol.~53,
  no.~4, p.~2046, 1996.

\bibitem{holmes2020impact}
A.~Holmes, S.~Johri, G.~G. Guerreschi, J.~S. Clarke, and A.~Y. Matsuura,
  ``Impact of qubit connectivity on quantum algorithm performance,'' {\em
  Quantum Science and Technology}, vol.~5, no.~2, p.~025009, 2020.

\bibitem{murali2019noise}
P.~Murali, J.~M. Baker, A.~Javadi-Abhari, F.~T. Chong, and M.~Martonosi,
  ``Noise-adaptive compiler mappings for noisy intermediate-scale quantum
  computers,'' in {\em Proceedings of the twenty-fourth international
  conference on architectural support for programming languages and operating
  systems}, pp.~1015--1029, 2019.

\bibitem{nishio2020extracting}
S.~Nishio, Y.~Pan, T.~Satoh, H.~Amano, and R.~V. Meter, ``Extracting success
  from ibm’s 20-qubit machines using error-aware compilation,'' {\em ACM
  Journal on Emerging Technologies in Computing Systems (JETC)}, vol.~16,
  no.~3, pp.~1--25, 2020.

\bibitem{gottesman1997stabilizer}
D.~Gottesman, {\em Stabilizer codes and quantum error correction}.
\newblock California Institute of Technology, 1997.

\bibitem{li2016hierarchical}
Y.~Li and S.~C. Benjamin, ``Hierarchical surface code for network quantum
  computing with modules of arbitrary size,'' {\em Physical Review A}, vol.~94,
  no.~4, p.~042303, 2016.

\bibitem{lameter2013numa}
C.~Lameter, ``Numa (non-uniform memory access): An overview: Numa becomes more
  common because memory controllers get close to execution units on
  microprocessors.,'' {\em Queue}, vol.~11, no.~7, pp.~40--51, 2013.

\end{thebibliography}

\end{document}